\documentclass[12pt]{iopart}

\usepackage{hypernat}

\usepackage{hyperref}

\hypersetup{pdfpagelabels=true,
	colorlinks=true,
	linkcolor=blue,
	anchorcolor=blue,
	menucolor=blue,
	urlcolor=blue,
	citecolor=blue,
	filecolor=blue,
	urlcolor=blue,
	pagecolor=blue,
	pdfauthor={Alberti et al.},
	pdftitle={Super-resolution microscopy of single atoms in optical lattices}}

\usepackage{color,graphicx}

\usepackage[utf8]{inputenc}
\usepackage[titletoc]{appendix}

\usepackage{booktabs}

\usepackage{iopams}   
\usepackage{txfonts}  
\usepackage{tabularx}

\usepackage[english]{babel}
\usepackage{cite} 
\usepackage{xspace} 

\newcommand{\figref}[2]{\hyperref[#1]{\ref{#1}(#2)}}
\newcommand{\figre}[1]{\hyperref[#1]{\ref{#1}}}
\renewcommand\textemdash{\leavevmode\unskip\kern0.8pt\rule[0.19\baselineskip]{8pt}{0.4pt}\kern1pt\ignorespaces}

\usepackage[separate-uncertainty=true]{siunitx}

\DeclareSIUnit\pixel{\ensuremath{\text{pixel}}}
\DeclareSIUnit\sqrtpixel{\ensuremath{\sqrt{\text{pixel}}}}
\DeclareSIUnit\micron{\micro\metre}
\DeclareSIUnit\mrad{\milli\rad}
\DeclareSIUnit\gauss{G}
\DeclareSIUnit\photon{\text{photon}}
\DeclareSIUnit\photoelectron{\text{ph.\hspace{2pt}e}${}^{-}$}
\DeclareSIUnit\electron{\text{e}^{-}}

\newif\ifusebibfile
\usebibfilefalse

\makeatletter
\renewcommand\l@section[2]{%
  \ifnum \c@tocdepth >\z@
    \addpenalty{\@secpenalty}%
    \addvspace{0.2em \@plus\p@}
    \setlength\@tempdima{1.5em}%
    \begingroup
      \parindent \z@ \rightskip \@pnumwidth
      \parfillskip -\@pnumwidth
      \leavevmode \bfseries
      \advance\leftskip\@tempdima
      \hskip -\leftskip
      #1\nobreak\hfil \nobreak\hbox to\@pnumwidth{\hss #2}\par
    \endgroup
  \fi}
\makeatother

\newcommand\sigmareadout{\ensuremath{\sigma_{\hspace{-1.5pt}\mathrm{readout}}}}
\newcommand\sigmaShotnoise{\ensuremath{\sigma_{\hspace{-1.5pt}\mathrm{fluo}}}}
\newcommand\sigmaStray{\ensuremath{\sigma_{\hspace{-1.5pt}\mathrm{stray}}}}

\newcommand\rms{RMS\xspace}

\newcommand\NAobj{\ensuremath{\mathrm{NA}_\textrm{\hspace{0pt}obj.~lens}}\hspace{-1pt}}

\newcommand\eqref[1]{(\ref{#1})}

\newcommand*\diff{\mathop{}\!\mathrm{d}} 

\newcommand\comb{\mathop{\mathrm{comb}}\nolimits}

\begin{document}
\selectlanguage{english}

\title[]{Super-resolution microscopy of single atoms in optical lattices
}

\author{Andrea~Alberti$^1$, Carsten Robens$^1$, Wolfgang~Alt$^1$, Stefan Brakhane$^1$, Micha{\l}~Karski$^1$, Ren{\'e}~Reimann$^1$, Artur~Widera$^2$, and Dieter~Meschede$^1$}

\address{$^1$ Institut f{\"u}r Angewandte Physik der Universit{\"a}t Bonn, Wegelerstr. 8, 53115 Bonn, Germany}
\address{$^2$ Fachbereich Physik und Landesforschungszentrum OPTIMAS, Gottlieb-Daimler-Str., 67663 Kaiserslautern, Germany}

\date{\today}


\begin{abstract}

We report on image processing techniques and experimental procedures to determine the lattice-site positions of single atoms in an optical lattice with high reliability, even for limited acquisition time or optical resolution.
Determining the positions of atoms beyond the diffraction limit relies on parametric deconvolution in close analogy to methods employed in super-resolution microscopy.
We develop a deconvolution method that makes effective use of the prior knowledge of the optical transfer function, noise properties, and discreteness of the optical lattice.
We show that accurate knowledge of the image formation process enables a dramatic improvement on the localization reliability.
{This allows us to demonstrate super-resolution of the atoms' position in closely packed ensembles} where the separation between particles cannot be directly optically resolved.
Furthermore, we demonstrate experimental methods to precisely reconstruct the point spread function with sub-pixel resolution from fluorescence images of single atoms, and we give a mathematical foundation thereof.
We also discuss discretized image sampling in pixel detectors and provide a quantitative model of noise sources in electron multiplying CCD cameras.
{The techniques developed here are not only beneficial to neutral atom experiments, but could also be employed to improve the localization precision of trapped ions for ultra precise force sensing.}

\end{abstract}

\vspace{2pc}
\vspace{28pt plus 10pt minus 18pt}
     \noindent{\small\it Keywords: {\rm fluorescence imaging of individual atoms, super-resolution imaging, optical lattices}\par}
		 		 
\maketitle
\tableofcontents
\section{Introduction}\label{sec:intro}

Detection and manipulation of individual atoms on neighboring sites of an optical lattice have attracted great interest in recent years for applications in quantum information processing~\cite{Zhang:2006,Lee:2007,Bloch:2008p128,Kar09,Karski:2010,Lee:2013,Wang:2015,Xia:2015}, quantum simulations~\cite{Bloch:2012jy,Bak09,She10,Karski2009a,Yamamoto:2015}, and very recently for studying strongly correlated Fermi systems at the single particle level~\cite{Haller:2015,Cheuk:2015,Parsons:2015,Omran:2015}.
Resolving atom positions with single-site resolution represents a technological challenge, since in optical lattices the distance between two lattice sites is on the order of the optical lattice wavelength.
In fact, experiments relying on atoms tunneling between lattice sites require short lattice constants since the tunneling rate decreases exponentially with larger intersite separation.

Previously, we demonstrated that the number of lattice sites between well-isolated atoms in a one-dimensional (1D) optical lattice can be resolved with high reliability even with objective lenses of moderate numerical aperture (NA) \cite{Kar09}.
Single-atom localization methods are employed in our laboratories ever since to measure the spatial probability distribution of atoms performing discrete-time quantum walks~\cite{Karski2009a, Genske2013, Robens2015}.
Recent experiments {in our laboratory} beyond single particle physics require resolving the position of each individual atom in small clusters at high filling factors, even when each lattice site is occupied~\cite{Robens:2016}.
By exploiting the discreteness of the atoms' positions in the lattice, we demonstrate in this manuscript new methods that enable resolving clusters of atoms with high reliability.
{Super-resolving a cluster of atoms constitutes a much bigger challenge than resolving the distance between exactly two atoms, as we originally demonstrated~\cite{Kar09}.}

{Besides presenting a conceptually simple introduction to super-resolved fluorescence imaging of atoms and to the related deconvolution problem, this work develops several new methods with respect to our original work \cite{Kar09}, which include: Wiener deconvolution of fluorescence images combined with a robust spectral-density-estimation algorithm for a first estimation of the atoms' positions (Sec.~\ref{sec:deconvolotion-process}), a weighted non-linear least squares estimation of positions accounting for the experimentally characterized noise (Sec.~\ref{sec:noise-model}), and inclusion of constraints of the atoms' positions on the periodic lattice (Sec.~\ref{sec:enhancing}), as well as an optimal algorithm for the iterative reconstruction of the line spread function of the imaging system (the analogue of the point spread function for 1D imaging) (Sec.~\ref{sec:determining-lsf}) with mathematical treatment of the convergence limit (App.~\ref{sec:math-convergence-lsf}), and a measurement of optical aberrations from single atom images (Sec.~\ref{sec:characterizing-the-lsf}).
}

In general, our methods for the analysis of fluorescence images are closely related to those employed in superresolution microscopy of biological structures~\cite{Huang2010,Thompson:2012,Pertsinidis:2010}, or in astronomy, where stars appear as point-like radiation sources~\cite{Starck:2006}:
Knowing the exact number of emitters in an observed region of interest allows us to determine the center position of each emitter with an uncertainty much smaller than the width of the point spread function (PSF) of the imaging system.
Super-resolution imaging relies on the precise knowledge of the properties of point-like atomic emitters trapped in an optical lattice as well as the detailed properties of background noise. 

{Light sources separated by less then an Abbe radius, $r_\text{A}=\lambda_\mathrm{f}/(2\mathrm{NA})$ ($\lambda_\mathrm{f}$ is the fluorescence radiation wavelength and $\text{NA}$ is the numerical aperture of the imaging system) form blurred, overlapping images, which cannot be optically distinguished and require super-resolution techniques to be resolved.}
In the past few years, {the Abbe diffraction limit has prompted ultracold atom experiments to develop objective lenses with larger numerical apertures in} the range of $0.6 {<} \mathrm{NA} {<} 0.8$ {to significantly enhance the optical resolution, thus {allowing single lattice sites to be optically resolved} \cite{Bak09,She10}.
These experiments have developed different solutions to the deconvolution problem: for example, linear least-squares fit of fluorescence intensities by a sum of reconstructed PSFs on a fixed lattice combined with threshold criterion \cite{Bak09,Parsons:2015,Yamamoto:2015,Omran:2015,Cheuk:2015}, deconvolution through a kernel function containing the PSF information (though not noise information) and threshold criterion \cite{Greif:2015}, deconvolution through a Gaussian kernel combined with a threshold criterion \cite{Miranda:2015}, deconvolution by fitting different configurations of occupancies on a fixed lattice (though without specifically mentioning methods to account for noise) \cite{She10}, or simply using a threshold criterion on the integrated fluorescence in the pixels corresponding to each lattice site \cite{Haller:2015}.}
A different approach based on electron microscopes has demonstrated even higher spatial resolution to resolve atoms in an optical lattice, though without reaching the sensitivity level needed for detecting single atoms yet~\cite{Ger08}.
%

{
However, experiments working under conditions of a low signal-to-noise ratio or employing a moderately large numerical aperture (as is our case $\text{NA}=0.23$) require methods that can extract the maximum physical information on the positions of atoms, especially when bunched in closely packed clusters.
The Fisher information matrix provides the mathematical instrument to identify the fundamental limit on the information that an estimator of positions can extract from a fluorescence image (Cramér-Rao information bound) \cite{Ober:2004}:
If such an estimator exists, than this estimator maximizes the likelihood function associated with the estimated quantity (i.e.\ positions).
In this sense, the maximum-likelihood estimator defines the gold standard for any image analysis technique.
As argued in Sec.~\ref{sec:deconvolotion-process}, we can approach this limit relying on a accurate knowledge of the line spread function and noise characteristics \cite{Abraham:2009}.
}
%
%
%

{Furthermore, we would like to remark that the techniques and results demonstrated in this work could have an impact even beyond neutral atom experiments.
For example, the majority of techniques for photoactivated localization microscopy have been optimized for the situation of fully separated fluorophore, while we here deal specifically with the opposite situation of bunched emitters (atoms) \cite{Small:2014}.
Our methods can also find application to improve the localization precision of trapped ions, where displacements of the equilibrium position are recorded to sense extremely tiny forces on the yoctonewton scale \cite{Biercuk:2010,Anetsberger:2010,Gloger:2015,Wong-Campos:2015}.
Recently, the same techniques presented in section~\ref{sec:characterizing-the-lsf} to quantitatively reconstruct optical aberrations have been employed, concurrently with this work, to characterize the imaging system of single trapped ions \cite{Wong-Campos:2015}.
}

\section{The deconvolution problem}\label{sec:deconvolution-problem}

\begin{figure}
    \centering
    \includegraphics[width=0.95\columnwidth]{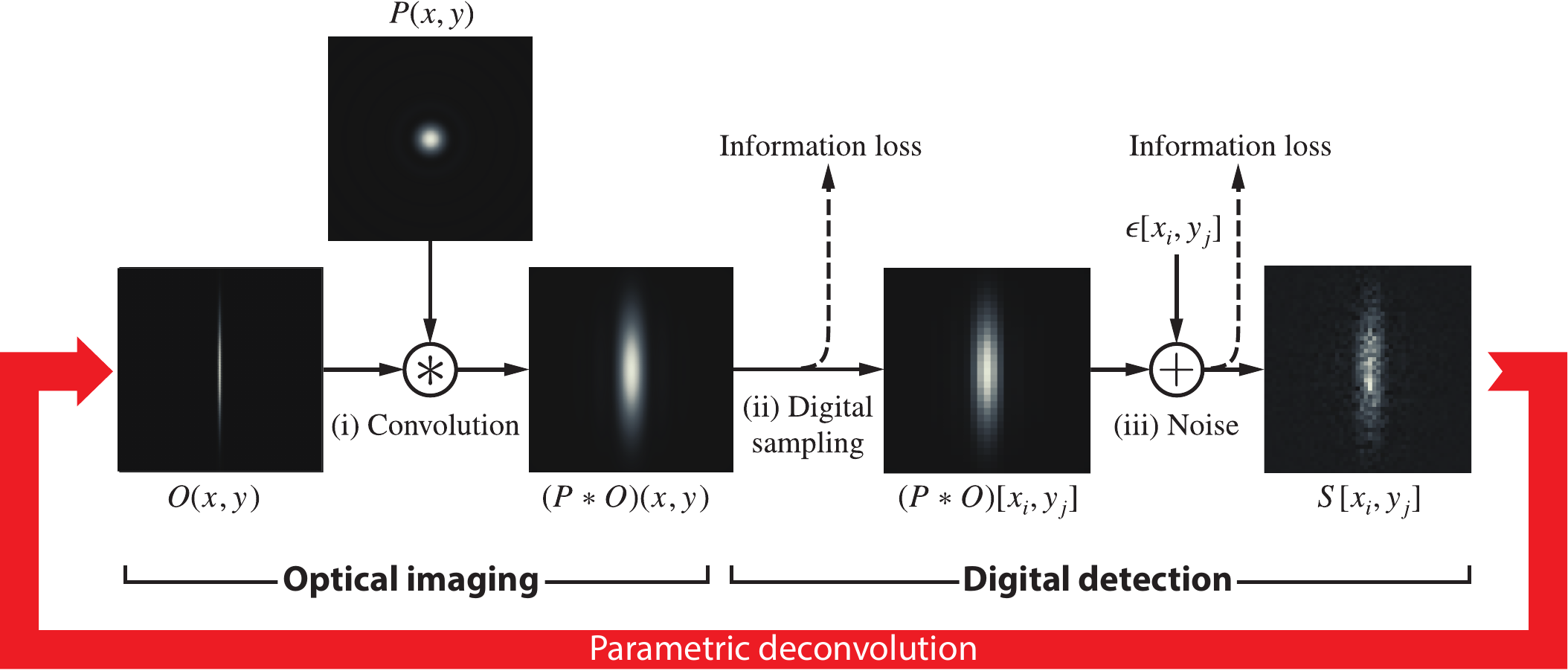}
    \caption{Schematic representation of image formation and information extraction.
    We retrieve the original atomic distribution $O(x,y)$ from the measured fluorescence image $S[x_i,y_j]$ by parametric deconvolution with the point spread function $P(x,y)$ of the imaging system. 
    Here we use the model assumption that atoms trapped in a 1D optical lattice are line-like radiation sources:
    Their motion is tightly confined along the longitudinal direction (horizontal direction in the images) and optically not resolved (see also~\sref{sec:atomic-sources}), while it is only loosely confined along the transverse direction (vertical).}
    \label{fig:image-formation}
\end{figure}

The steps involved in the image formation \textemdash from the point-like atomic radiation sources to
the final image on the digital camera \textemdash are schematically depicted in \fref{fig:image-formation} and summarized below:
\begin{enumerate}

\item Optical diffraction by the imaging microscope transforms all
radiation sources into blurred spatial distributions. This process
can be mathematically expressed as the convolution of the original
distribution $O_{}(x,y)$ with the point spread function (PSF)
$P_{}(x,y)$, which represents the characteristic intensity distribution of an
ideal point source recorded by the imaging system. In case of a
system whose optical response is translationally invariant
(isoplanatic behavior), the fluorescence image distribution reads
\begin{eqnarray}
	I_\mathrm{fluo}^{}(x,y) &=&(P_{}^{}\ast O_{}^{})(x,y) = \int_{-\infty}^{\infty}\int_{-\infty}^{\infty}\hspace{-2pt} P_{}^{}(x-u,y-v)\hspace{2pt}O_{}^{}(u,v)\,\diff u \diff v\,.\label{eq:2d-continuous-imaging-equation} 
\end{eqnarray}

In case of a hard circular aperture, for example, the PSF of diffraction-limited aberration-free optical system is the well-known Airy disc pattern~\cite{Goo96}, whose first minimum corresponds to $1.22\,r_\mathrm{A}$, with $r_\text{A}$ being the Abbe radius defined above.

\item A CCD detector samples and digitizes the image distribution $I(x,y)\mapsto I[x_{i}^{},y_{j}^{}]$, where $x_{i}^{}$ and $y_{j}^{}$ denote the integer-valued horizontal and vertical positions of a sampling bin and the squared brackets in our notation distinguish discrete from continuous distributions because in general ${I_{}^{}(x_{}^{},y_{}^{})\not= I_{}^{}[x_{i}^{},y_{j}^{}]}$.
In fact, the discrete and continuous distribution are mathematically related through
\begin{equation}
	\label{eq:sampling} \hspace{-1.5cm}I[x_i,y_j]=\hspace{-7pt}\sum_{n,m=-\infty}^\infty\hspace{-3pt}\mathcal{R}_\text{s}(x_i-n\,\Delta_\mathrm{s},y_j-m\,\Delta_\mathrm{s})\,\int_{-\infty}^\infty\hspace{-2pt} \mathcal{R}_\mathrm{p}(n\,\Delta_\text{s}-u,m\,\Delta_\text{s}-v)\,I(u,v)\,\diff u \diff v 
\end{equation}
where $\Delta_\mathrm{s}$ represents the sampling spacing in the object plane, $\mathcal{R}_\mathrm{p}(u,v)$ is the CCD pixel rectangular function equal to $1/\Delta_\mathrm{p}^2$ for $u$ and $v$ in the interval $[-\Delta_\mathrm{p}/2,\Delta_\mathrm{p}/2]$ and zero outside, with $\Delta_\mathrm{p}<\Delta_\mathrm{s}$ being the size of the pixel projected onto the object plane (today's CCD detector have $\Delta_\text{p}\sim\Delta_\text{s}$).
Likewise, $\mathcal{R}_\mathrm{s}(u,v)$ is the sampling rectangular function equal to one for both $u$ and $v$ in the interval $[-\Delta_\mathrm{s}/2,\Delta_\mathrm{s}/2]$ and zero elsewhere.
Equation~(\ref{eq:sampling}) represents the convolution of the continuous intensity distribution with the pixel response function $\mathcal{R}_\text{p}(u,v)$ (digitization), which is multiplied by the 2D $\comb$ function with spacing $\Delta_\mathrm{s}$ (discrete sampling)~\cite{Wittenstein:1982,Park:1984,Boreman2001}.
In order to prevent information loss by discrete sampling, the Nyquist-Shannon sampling theorem shows that the PSF \textemdash or, more precisely, the Abbe radius \textemdash must be imaged onto more than two CCD pixels, i.e., $r_\mathrm{A}>2\,\Delta_\mathrm{s}$~\cite{Gaskill1978,Pawley:2006}.

\item Physical information contained in the recorded signal $S[x_i,y_j]$ is partially lost due to diverse noise sources affecting the image formation process.
The effect of noise sources (see~Tab.~\ref{tab:noise-contributions} for a complete list) is taken into accounted through an additive stochastic noise term $\epsilon[x_{i}^{}, y_{j}^{}]$, which is added to the fluorescence intensity distribution:
\begin{equation}\label{eq:2d-discrete-imaging-equation}
    S_{}^{}[x_{i}^{},y_{j}^{}]=I_{\mathrm{fluo}}[x_i,y_j]+\epsilon{}^{}[x_{i}^{},y_{j}^{}]=
    (P_{}^{}\ast O_{}^{})[x_{i}^{},y_{j}^{}]+\epsilon_{}^{}[x_{i}^{},y_{j}^{}]\,.
\end{equation}
Here we assumed that the homogeneous background (by digitization offset, stray light, and dark currents discussed in Sec.~\ref{sec:noise-characterization}) is subtracted from the signal so that the average value of the noise vanishes, $\langle\epsilon[x_{i}^{}, y_{j}^{}]\rangle = 0$. 
\end{enumerate}

\noindent{}In order to retrieve the original information $O(x,y)$ from $S[x_i,y_j]$, we need to invert equation~(\ref{eq:2d-discrete-imaging-equation}) through deconvolution.
However, deconvolution problems are in general ill-conditioned, especially in the presence of noise. A physical model assumption \textemdash known as regularization \textemdash must be employed to constrain the solutions.
For example, atoms which are strongly confined in an optical lattice are modeled as identical, isolated emitters characterized by only two parameters: positions and fluorescence intensity.
Numerous deconvolution strategies exist in the literature, differing in their effectiveness in constraining solutions and in the computational resources required~\cite{Sta02}.
{Our specific parametric deconvolution approach mainly relies on a maximum-likelihood estimation constrained on a discrete lattice, for which a first estimation of the atoms’ positions is provided by the so-called MUSIC (multiple signal classification) algorithm (see Sec.~\ref{sec:parametric-deconvolution}).}

\section{The detection apparatus}\label{sec:detection-apparatus}

\begin{figure}[t!]
    \centering
    \includegraphics[width=0.9\columnwidth]{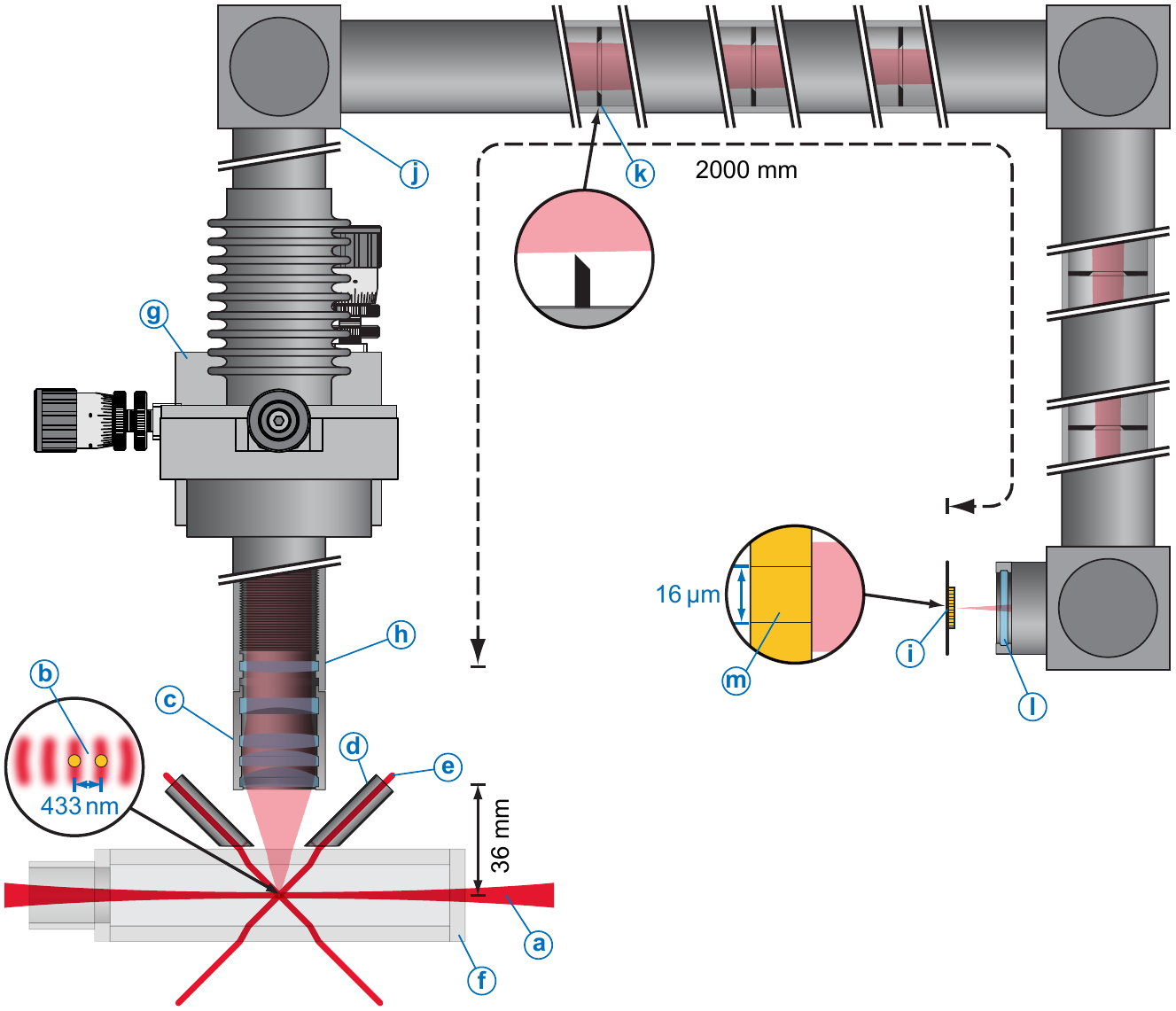}
    \caption{%
Illustration of the single-atom microscope.
A 1D optical lattice produced by two counterpropagating laser beams (\textbf{a}) traps atoms (\textbf{b}) in the object plane of a infinity-corrected microscope objective (\textbf{c}).
Beam tubes (\textbf{d}) block stray light and reflections of molasses beams (\textbf{e}) off the glass cell (\textbf{f}).
A three-axis translation stage (\textbf{g}) allows precise alignment of the microscope objective.
A tube lens (\textbf{h}) focuses the image onto an EMCCD detector (\textbf{i}).
Tubes and cube-mounted turning mirrors (\textbf{j}) bridge the distance between the tube lens and the detector, while several built-in knife-edge apertures (\textbf{k}) and a narrow-band optical filter (\textbf{l}) further suppress remaining stray light.
The separation between two adjacent lattice sites in the object plane is imaged to $\SI{24}{\micron}$ in the image plane, which amounts to about $\num{1.5}$ CCD pixels (\textbf{m}).}
\label{fig:imaging-setup}
\end{figure}

\subsection{The optical microscope}\label{sec:imaging-setup}
The imaging system depicted in~\fref{fig:imaging-setup} realizes an infinity-corrected microscope:
The fluorescence light emitted by the atoms at $\lambda_\text{f}=\SI{852}{\nano\meter}$ is collimated by a diffraction-limited objective lens (effective focal length ${f_1=\SI{36}{\milli\meter}}$)~\cite{Alt02} and imaged onto an EMCCD detector by a plano-convex tube lens (focal length $f_2=\SI{2}{\meter}$).
The magnification of the microscope is $f_2/f_1 \sim \num{55}$, so that the Abbe radius ($r_\mathrm{A}= \SI{1.9}{\micro\meter}$) of the point spread function is imaged over $>\num{6}$ pixels of the CCD camera, thus fulfilling the requirement by the Nyquist-Shannon sampling theorem to avoid information loss \cite{Gaskill1978,Pawley:2006}.
%

We assembled the microscope objective from four off-the-shelf spherical lenses, which are stacked into a one-inch aluminum holder.
By design the objective lens operates at the diffraction limit with a numerical aperture of $\NAobj=0.29$.
The objective lens was characterized with a shear-plate interferometer resulting in a peak-valley wavefront distortion of less than $\lambda_\mathrm{f}/4$ over $90\%$ of the clear aperture fulfilling Rayleigh's quarter wavelength rule~\cite{BornWolf}.
The long working distance ($\SI{36.5}{\milli\meter}$) allows the objective lens to be mounted outside of the vacuum sufficiently far away from the vacuum cell to prevent reflected light from molasses laser beams from reaching the camera.
The microscope objective is mounted on a three-axis translation stage and connected through blackened tubes to the EMCCD detector.
Inside the tubes five sooted knife-edge apertures are placed with gradually decreasing inner diameters to block stray light.
To further suppress the remaining stray light a narrow-band ($\SI{3}{\nano\meter}$ FWHM) optical filter with a transmission of $98\%$ at the wavelength $\lambda_{{\mathrm{f}}}$ is placed in front of the EMCCD detector.
Over the past few years, single aspheric lenses with long working distance (a few $\si{cm}$) have become widely available, and their utilization represents a good alternative to build an imaging system with a moderate numerical aperture ($0.2 < \rm NA < 0.5$).

\subsection{Localization of trapped atoms with small number of photons}\label{sec:small-light-signals}
According to Abbe's diffraction limit, the optical resolution of our imaging system is {$r_\text{A}=\SI{1.9}{\micro\meter}$}. However, to achieve single-site resolution we need to extract the position of single trapped atoms with an uncertainty smaller than  the lattice spacing $a$ ($\SI{433}{\nano\meter}$).
In analogy to super-resolution imaging in biological systems, we can determine the position of our atoms beyond the optical resolution by precisely knowing its point spread function and the underlying noise.
{Following reference~\citenum{Bobroff:1986}, in one dimension the localization precision of the fluorescence peak produced by a single atom can be estimated by
\begin{equation}\label{eqn:localozation-precision}
    (\Delta x)^2 = \frac{\mathrm{\rms}_\mathrm{PSF}^2+{\Delta_\text{p}^2}/{12}}{N}+\frac{4\sqrt{\pi}\cdot \mathrm{\rms}_\mathrm{PSF}^3\,\sigma_\text{b}^2\, {n_\perp}}{\Delta_\text{p}\hspace{1pt}N^2}\,,
\end{equation}
where it is assumed that the fluorescence signal is integrated over $n_\perp$ pixels in the direction transverse to the lattice, and that $\mathrm{\rms}_\text{PSF}$ is the \rms width of a Gaussian point spread function, $\Delta_\text{p}$ is the size of a camera pixel in the object plane, $N$ is the average number of recorded photons per atom, and $\sigma_\text{b}$ is the \rms background noise (see Eq.~(\ref{eq:emccd-noise-model}) in Sec.~\ref{sec:noise-characterization}).
In the literature, extensions of the result in equation~(\ref{eqn:localozation-precision}) can be found for two dimensions \cite{Thompson:2002} and, using the statistical theory based on the Fisher information matrix, for a generic disc point-spread function (e.g., Airy disc) \cite{Ober:2004}.
The Fisher information approach, which for a Gaussian point spread function yields exactly equation~(\ref{eqn:localozation-precision}), produces the fundamental theoretical localization limit that can be attained (Cramér-Rao information bound).
Note also that the localization precision in equation~(\ref{eqn:localozation-precision}) concerns only a single localized emitter, which is the case, for example, of an isolated fluorophore in photoactivated localization microscopy or of a very sparsely filled optical lattice.
Section~\ref{sec:parametric-deconvolution} is in particular concerned to super-resolve the position of emitters (atoms) clustered in small ensembles, which constitutes a significantly more demanding task.
In addition, it should be noticed that, when employing an electron multiplying CCD camera (as is the case of the present work), a factor 2 must to be added in front of $\mathrm{\rms}_\mathrm{PSF}^2$ in equation (\ref{eqn:localozation-precision}) to account for the effectively halved quantum efficiency due to the electron multiplying excess noise (see Sec.~\ref{sec:noise-characterization}) \cite{Quan:2010}.
}

In the following, we intend to give an estimate of the localization precision of our imaging system based on equation~(\ref{eqn:localozation-precision}): The $\mathrm{\rms}_\text{PSF}$ of our imaging system is $\sim\SI{1.5}{\micro\meter}$ (see Sec.~\ref{sec:atomic-sources}) and the parameter $\Delta_\text{p}$ can be calculated by dividing the pixel size ($\SI{16}{\micro\meter}$) by the magnification ($\sim\num{55}$, see Sec.~\ref{sec:imaging-setup}). The number of photoelectrons ($\si{\photoelectron}$) recorded on the EMCCD sensor per single atom can be estimated by knowing the photon scattering rate, the solid angle of the microscope objective into which photons are emitted, and the exposure time.
Atoms illuminated with nearly resonant light at $\lambda_\mathrm{f}$ emit photons at the maximal rate of $\Gamma/2$ for strong saturation, with $\Gamma\sim 2\pi \times \SI{5}{\mega\hertz}$ being the radiative decay rate for cesium.
However, to prevent atoms from hopping along the lattice during imaging, the saturation parameter is typically chosen much smaller~\cite{Miroshnychenko:2006,Fuhrmanek:2011} ($s\sim 0.01$), which reduces the scattering rate by a factor of 10 or more~\cite{Hume:2013}.
The solid angle directly depends on the NA of the imaging system according to the formula $\Omega/4\pi=(1-\sqrt{1-NA^2})/2\sim \SI{1}{\percent}$. By additionally taking into account the finite quantum efficiency of the CCD camera $\mathrm{QE}(\lambda_\mathrm{f})\sim \SI{30}{\percent}$ (see Sec.~\ref{the-emccd-detector}) as well as photon losses ($\sim\SI{6}{\percent}$) due to both reflections from optical surfaces (e.g.\ the vacuum glass cell) and the transmission of the narrow-band optical filter (see Sec.~\ref{sec:imaging-setup}), we expect to detect about $\SI{1000}{\photoelectron}$ per atom for a single fluorescence image with an exposure time of $T=\SI{1}{\second}$.
For comparison, in our experiments we record about $\SI{1300}{\photoelectron\,s^{-1}}$ per atom as discussed in~\sref{sec:image-segmentation}.
The measured background-noise distribution, which is analyzed in~\sref{sec:noise-model}, has a \rms width $\sigma_\text{b}$ of about $\SI{0.6}{\photoelectron}$ per camera pixel.
Since we integrate the fluorescence images along the direction transverse to the 1D optical lattice (see Sec.~\ref{sec:determining-lsf}), the variance of the background noise $\sigma_b^2$ is multiplied by the number of transverse pixels ${n_{\perp}}$ (typically $n_\perp\sim40$).
Hence, based on equation~(\ref{eqn:localozation-precision}) we expect a localization precision of {$\Delta x \sim \SI{60}{\nano\meter}$}, which is sufficiently smaller than the separation between two lattice sites.
By using longer exposure times it is possible to improve the resolution even further, however, at the cost of decreasing the duty cycle and increasing the probability for atoms to either hop to adjacent lattice site or to be lost because of heating and background gas collisions.
{Moreover, a slow drift of the entire lattice with respect to the imaging system ($\leq 20\,\mathrm{nm}/\mathrm{s}$ \cite{Karski:2010}) is responsible for the existence of an optimal exposure time (estimated around $\SI{2}{\second}$ for our system) beyond which the localization precision deteriorates instead of improving, if the lattice drift is not suitably tracked and accounted for.
Such drifts are especially notable in case of imaging systems with very high optical resolution, as recently demonstrated through the measurement of the Allan variance associated with the position uncertainty of trapped ions  \cite{Wong-Campos:2015}.}

\subsection{The EMCCD detector}\label{the-emccd-detector}

In our experiment, the fluorescence signal of the atoms is detected using an electron multiplying CCD (EMCCD) camera (Andor iXon DV897DCS-FI), whose read-out noise is more than one order of magnitude smaller compared to that of scientific-grade CCD sensors.
In fact, scientific-grade CCD sensors are at present limited by a background noise $\sigma_\text{b}>\SI{6}{\photoelectron}$ per pixel.
The increased noise would deteriorate the localization precision estimated for our system by a factor of 6  (see Eq.~(\ref{eqn:localozation-precision})), therefore preventing reliable single-site resolution.
To detect signals of few photons, such as fluorescence of single atoms, alternative types of imaging sensors exist, which include intensified CCD sensors and CMOS sensors.
In appendix~\ref{sec:detector_comparison} we provide a review of sensors suited for few-photon-signal detection, and discuss technical noise sources inherent the different technologies are discussed.
Our EMCCD camera employs a front-illuminated, frame-transfer, buried channel CCD sensor (L3Vision CCD97) containing ${512\times 512}$ active pixels with a pixel size of ${16\times 16\,\mu\mathrm{m}^2}$.
A measurement comparing front- with back-illuminated sensors is given in appendix~\ref{sec:comparing-detectors}.
The quantum efficiency at the imaging wavelength ${\lambda_{\mathrm{f}}}$ with the EMCCD sensor cooled to $\SI{-70}{\celsius}$ is $\mathrm{QE}(\lambda_{\mathrm{f}})\sim \SI{30}{\percent}$.
It should be noted that the efficiency decreases at lower temperatures for wavelengths $>\SI{800}{\nano\meter}$ due to a temperature dependence of the silicon band gap~\cite{Sperlich:2013}.

In EMCCD sensors, suppression of read-out noise is achieved through the serial electron multiplying (EM) register, which amplifies charges by impact ionization at each shift step, similar to a staircase avalanche photodiode (see App.~\ref{sec:detector_comparison}).
The EM register of our camera comprises ${N=536}$ shift steps.
Even though the probability of impact ionization at each individual step is only ${p_{\mathrm{imp}}^{}\sim 1.5\%}$, due to the large number of steps, the mean gain of the cascaded multiplication process ${g=(1+p_{\mathrm{imp}}^{})_{}^{N}}$ can reach values well above $1000$.
The effect of its stochastic nature on the detection noise is discussed in section~\ref{sec:noise-characterization}.

\section{The atomic sources}\label{sec:atomic-sources}
\begin{figure}
    \centering
    \includegraphics[width=0.87\columnwidth]{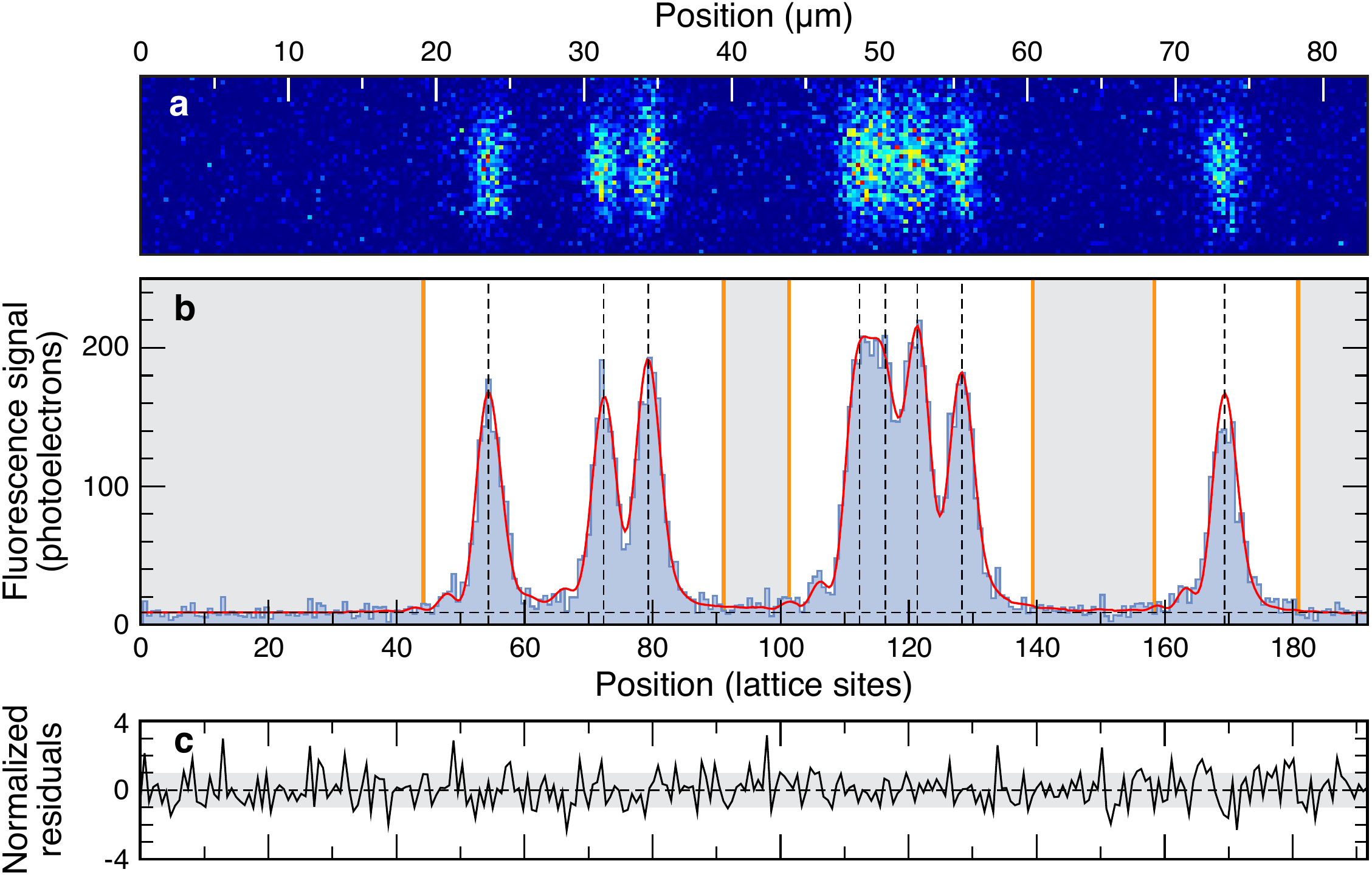}
    \caption{\label{fig:image-segmentation}
	(\textbf{a})
	Image of atoms in a 1D optical lattice acquired with a $\SI{1}{\second}$ exposure time.
(\textbf{b}) The corresponding integrated intensity distribution $I[x_i]$.
The image is subdivided into regions of interest (white regions) and regions with no fluorescence signal (gray regions), which are used to determine the constant background baseline (dashed horizontal line).
The solid red line shows the result of the parametric deconvolution, where the vertical dashed lines show the positions of the atoms constrained on a periodic lattice.
The distance of the atoms from the leftmost one are 18, 25, 58, 62, 67, 74, 115 lattice sites.
(\textbf{c}) Normalized residuals between the integrated fluorescence signal and the fitted model, resulting in a reduced $\chi^2=\num{0.835}$.
}
\end{figure}

To acquire fluorescence images with the detection apparatus described in section~\ref{sec:detection-apparatus} we trap atoms in a deep lattice potential with lattice constant $a=\SI{433}{\nano\meter}$ and illuminate them with a three-dimensional optical molasses.
The saturation parameter of the optical molasses and the lattice trap depth (${U/k_{\mathrm{B}}=0.4\,\mathrm{mK}}$) are chosen as such to prevent atoms from hopping along the lattice direction.
Figure~\figref{fig:image-segmentation}{a} exemplarily shows a fluorescence image of eight trapped atoms, which are loaded into a 1D optical lattice in stochastic positions and subsequently imaged with an illumination time of $\SI{1}{\second}$.
The intensity distribution for each atom exhibits a characteristic elliptical shape elongated along the radial direction of the optical lattice with an aspect ratio of about 6:1 (FWHM along the axial direction of $\SI{1}{\micro\meter}$).
The elongated shape originates from the thermal motion of trapped atoms ($\sim\SI{30}{\micro\kelvin}$ by sub-Doppler cooling in the optical molasses) in the radial direction, along which the confinement is weaker.
Along the lattice direction, instead, trapped atoms can be regarded as localized point sources with a Dirac-delta longitudinal distribution
\begin{equation}
	\label{eq:diracDist}
	O(x,y)= O(y)\, \delta(x)
\end{equation}
with a certain radial intensity distribution $O(y)$, because the extent of the axial thermal motion (FWHM $\sim \SI{60}{nm}$) as well as the drift of the optical lattice ($\leq 20\,\mathrm{nm}/\mathrm{s}$ \cite{Karski:2010}) is one order of magnitude smaller than the optical resolution.
Being primarily interested in extracting the precise position of atoms along the optical lattice,
we integrate the acquired images along the radial direction ($I[x_i] = \sum_{j} I[x_i,y_j]$) as depicted in figure~\figref{fig:image-segmentation}{b}, which reduces the deconvolution problem to a one-dimensional one.
The continuous curve overlapped with the integrated fluorescence signal shows the end result of the parametric deconvolution problem presented in section~\ref{sec:parametric-deconvolution}, which yields the atoms' positions with single lattice-site precision.
Figure~\figref{fig:image-segmentation}{c} shows the residuals between the reconstructed distribution and the measured signal, normalized to the expected noise strength.
The uniform distribution of residuals with \rms spread around one attests the quality of the parametric deconvolution, which is ensured by an accurate knowledge of the LSF of the imaging system as well as of the noise model.
The methods to reconstruct the LSF are presented in following section~\ref{sec:determining-lsf}, while the physical noise model is presented in section~\ref{sec:modelling-noise-sources} and the parametric deconvolution process is illustrated in the last section~\ref{sec:parametric-deconvolution}.

\subsection{Reconstructing the line spread function with sub-pixel resolution}\label{sec:determining-lsf}

One key element to achieve a resolution beyond the diffraction limit is the accurate knowledge of the response of our imaging apparatus.
More precisely, it is important for the parametric deconvolution problem to know exactly the imaged fluorescence intensity distribution of a single illuminated atom.
The importance of an accurate knowledge of this distribution is quantitatively demonstrated in section~\ref{sec:enhancing}.
In a 1D optical lattice, the problem of reconstructing the positions of atoms can be reduced to one dimension by exploiting the factorized form of the single-atom distribution in equation~(\ref{eq:diracDist}).
In fact, a single atom positioned at $x=0$ yields (see Eq.~(\ref{eq:2d-continuous-imaging-equation})) a fluorescence distribution that integrated along the radial direction reads
\begin{equation}
\int_{-\infty}^\infty I_\text{fluo}(x,y)\,\mathrm{d}y =L(x)  \int_{-\infty}^\infty O(y)\,\mathrm{d}y\,,
\end{equation}
where \raisebox{0pt}[0pt][0pt]{$L(x)=\hspace{-1pt}\int_{-\infty}^\infty P(x,y)\,\mathrm{d}y$} represents the so-called line spread function.
As argued in section~\ref{sec:deconvolution-problem}, the response function required in the deconvolution problem is the convolution of the optical line spread function $L(x)$ with the 1D CCD pixel function $\mathcal{R}_\textrm{p}(x)$~\cite{anderson2000},
\begin{equation}\label{eqn:ideal-LSF-reconstruct-convergence}
L_\mathrm{CCD}(x)=(\mathcal{R}_\textrm{p}\ast L)(x)\,.
\end{equation}
In the following we present our method to reconstruct the $L_\mathrm{CCD}$ function with increased signal-to-noise ratio and sub-pixel resolution, which is based on superimposing multiple intensity distributions of sufficiently isolated atoms (for example, the rightmost atom in Fig.~\figref{fig:image-segmentation}{b}).
The superimposing process is generally referred to as image registration in digital signal processing.

We make use of a recursive algorithm to process single-atom images, whose end result should ideally converge to $L_\text{CCD}$ in equation~(\ref{eqn:ideal-LSF-reconstruct-convergence}).
The algorithm is composed of a preparatory procedure followed by an iterative one.
The first step of the preparatory procedure consists in identifying those regions of interest containing exactly one atom well separated from other atoms by several Abbe radii (typically 10) in order to allow us not only to reconstruct the central peak of the LSF but also the wings containing the diffraction fringes.
In the next step, we apply a Fourier filter to each single-atom image to remove high-spatial-frequency noise.
The filter utilizes the fact that every optical system with a hard aperture has a cutoff in the optical transfer function (OTF), defined as the Fourier transform of $L_\mathrm{CCD}$, exactly at the Abbe frequency $1/r_\text{A}=2\mathrm{NA}/\lambda_\mathrm{f}$.
After discrete Fourier transformation (DFT) of the integrated intensity distributions, the filter sets the amplitude of all frequencies beyond the Abbe cut-off (typically $>\num{1.2}/r_\text{A}$ to reduce Fourier artifacts) to zero because these frequencies components do not carry physical information ($\text{OTF}=0$ in this region).
The effect of Fourier filtering is significant for our imaging system because the Abbe frequency is three times smaller than the Nyquist frequency of $\SI{0.5}{\pixel^{-1}}$ \textemdash the frequency up to which noise appears if not filtered out.
The last step of the preparatory procedure to reconstruct the LSF consists in interpolating the noise-filtered single-atom distributions with sub-pixel resolution, which allows us to reposition them in the subsequent iterative procedure with high precision.
Because of the finite bandwidth of the OTF, the integrated fluorescence signal can be interpolated with an arbitrary spatial resolution using the Whittaker–Shannon interpolation formula:
We extend the DFT fluorescence distribution in Fourier space beyond the Abbe cut-off with zero values (zero padding), so that the number of points in the Fourier space is increased by an integer factor $s$ with respect to the original number.
The inverse DFT of the zero-padded signal results in an upsampled distribution, where the width of a sub-pixel is equal $1/s$ of the original pixel's width.
The size of the sub-pixel is chosen smaller than the estimated localization precision (typically $s=8$ so that $\SI{1/8}{\pixel}=\SI{37}{\nano\meter}<\Delta x$, see~Sec.~\ref{sec:small-light-signals}).
An alternative yet equivalent application of the Whittaker–Shannon interpolation formula operates directly in position space by convolving the spatial distribution with a sinc function.

The iterative part of the reconstruction algorithm consists chiefly of two steps.
In the first one, we obtain the position of each atom by a non-linear least squares fit of the model distribution  $L_\text{CCD}$ to the recorded fluorescence signal (see Sec.~\ref{sec:parametric-deconvolution} for more details).
The precise (unrounded) value of the atom position is used to shift and align all noise-filtered sub-pixel-interpolated intensity distributions.
Hence, superimposing all images gives a reconstruction of the fluorescence distribution of a single atom with a signal-to-noise ratio enhanced by a factor $\sqrt{N_\text{at}}$, where $N_\text{at}$ is the number of superimposed single atoms (typically a few hundreds).
The reconstructed distribution $L_\mathrm{guess}$ provides us with a new estimate of $L_\text{CCD}$.
The iterative algorithm stops when no change is observed (typically after 5 to 10 iterations).
For the first iteration, we use a Gaussian function to determine the position of single atoms in case no LSF function is \emph{a priori} known.

\begin{figure}[t]
    \centering
    \includegraphics[width=1\columnwidth]{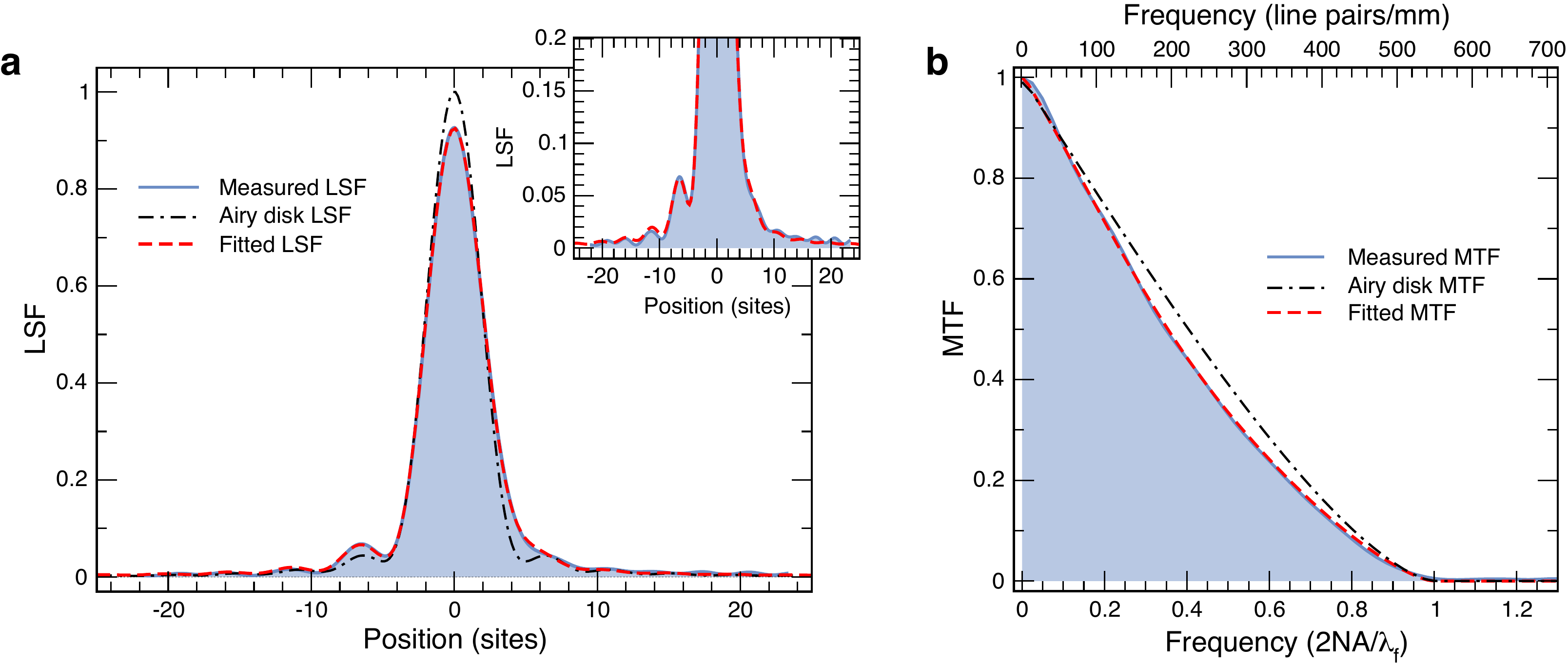}
    \caption{\label{fig:reconstructed-LSF-MTF}
(\textbf{a}) The solid blue line shows the reconstructed LSF from more than 200 single-atom images, the dash-dotted black line shows the ideal, diffraction-limited LSF derived from an Airy disk with $\mathrm{NA} = 0.228$, and the dashed red line represents the fitted model based on a wavefront expansion in Zernike polynomials.
The dash-dotted curve is normalized to have a maximum value of 1, while the other two curves are normalized to the same area of the dash-dotted one.
(\textbf{b}) Corresponding modulation transfer functions. All three curves show the hard cut-off at the Abbe frequency $1/r_A$.
}\label{fig:lsf-mtf}
\end{figure}

A mathematical derivation (see App.~\ref{sec:math-convergence-lsf}) shows that this algorithm converges to 
\begin{equation}\label{eqn:LSF-reconstruct-convergence}
L_\mathrm{guess}(x)=(\mathcal{R}_{x}\ast\mathcal{R}_\textrm{sp}\ast\mathcal{R}_\textrm{p}\ast L)(x)\,,
\end{equation}
instead of the desired expression in equation~(\ref{eqn:ideal-LSF-reconstruct-convergence}), where $\mathcal{R}_x$ is the probability distribution of the non-linear least squares estimator of the single-atom position for an atom ideally positioned in the origin $x=0$ (with a RMS width $\Delta x\sim \SI{60}{\nano\meter}$, see Sec.~\ref{sec:small-light-signals}), and $\mathcal{R}_\textrm{sp}$ the sub-pixel function equivalent to the pixel function $\mathcal{R}_\textrm{p}$ but $s$ times narrower.
Because the ``blurring'' effect by both additional convolutions in equation~(\ref{eqn:LSF-reconstruct-convergence}) is on the order of a few tens of nanometers, we conclude that $L_\mathrm{guess}(x) \sim (\mathcal{R}_\textrm{p}\ast L)(x)$ to a good approximation.
For precise reconstruction of the $L_\text{CCD}$ function, equation~(\ref{eqn:LSF-reconstruct-convergence}) shows that it is advantageous to increase the illumination time in order to decrease $\Delta x$.
The reconstructed LSF and the related modulation transfer function ($\text{MTF}=|\text{OTF}|$) are displayed in figure~\ref{fig:reconstructed-LSF-MTF} and analyzed in detail in the following section.

\subsection{Analysis of the reconstructed line spread function}\label{sec:characterizing-the-lsf}
Besides its importance to retrieve the atoms' positions with the maximum localization precision (see Sec.~\ref{sec:enhancing}), the line spread function contains valuable information about the performance of the optical system.
Since the influence of $\mathcal{R}_\text{p}$ in equation~(\ref{eqn:ideal-LSF-reconstruct-convergence}) is small (ensured by the Nyquist-Shannon condition $r_\text{A}>2\Delta_\text{s}$), the optical line spread function $L$ is well approximated by $L_\text{CCD}$.
Figure~\figref{fig:reconstructed-LSF-MTF}{a} shows the reconstructed LSF obtained with the algorithm outlined in the foregoing section.
In case of an aberration-free imaging system, the point spread function is described by the well-known Airy disk, whose corresponding LSF is displayed for comparison in the same figure.
Besides an overall agreement, the reconstructed LSF exhibits a lower maximum and a distinct asymmetry such that the higher-order diffraction peaks are only visible on the left-hand side.
These differences arise from wavefront distortion caused by optical aberration.
Mathematically, the point spread function is defined by computing the modulus square of the Fourier transform of the electric field (wavefront) at the pupil (Fraunhofer diffraction).
The wavefront contains all information about optical aberrations and can be expressed in the basis of Zernike polynomials \cite{BornWolf}.
To gain insight into the nature and amount of the optical aberrations affecting our optical system, we fitted to the reconstructed LSF the one obtained from a wavefront expansion in low-order Zernike polynomials up to spherical aberration.
The fitted LSF is displayed in the same figure, demonstrating a remarkable agreement with the experimental curve.
The fit analysis reveals that the leading aberration contribution arises from astigmatism.
The detailed list of Zernike coefficients is given in Tab.~\ref{tab:Zernike}.
Combining all contributions in the table yields an overall \rms wavefront error of $\sim\lambda/17$ (whereas the peak-valley deviation is $\lambda/3$), which corresponds to a Strehl ratio of $\num{0.87}$ defined as the ratio between the maxima of the measured PSF and the ideal one.
Note that the Strehl ratio, in general, differs from the ratio obtained analogously for the 1D LSF (about 0.92, see Fig.~\ref{fig:lsf-mtf}).
In addition, the wavefront analysis gives an estimate of the actual numerical aperture of the optical system, $\text{NA}=0.228(3)$.
The deviation between the estimated numerical aperture and the one of the objective lens design ($\mathrm{NA} = 0.29$) is most likely caused by clipping at the knife-edge apertures along the imaging path, see figure~\ref{fig:imaging-setup}.
{Concurrently with this work, a similar wavefront analysis based on Zernike polynomials has been carried out to characterize the aberrations affecting two-dimensional fluorescence images of single trapped ions \cite{Wong-Campos:2015}.}

\begin{table}[t]
	\caption{\label{tab:Zernike}Result of the wavefront fitting to the measured LSF expressed in terms of low-order Zernike polynomials. The overall wavefront distortion is obtained by adding the different contributions in quadrature. 1D fitting of our model to the LSF cannot prevent a certain ambiguity on the identification of wavefront distortion angles (not displayed).}
    \centering
	\vspace{2mm}
	\begin{tabular}{m{4.5cm}ccccc}
		\toprule\ns {} & Defocus & Astigmatism & Coma & Trefoil  & Spherical \\\ns
		\midrule
		Orders (radial, azimuthal)  & (2,0) & (2,2) & (3,1) & (3,3) & (4,0) \\
		RMS wavefront
		distortion \newline{} \raisebox{3pt}[0pt][0pt]{($\lambda$ units)} & $0.016(2)$ & $0.048(2)$ & $-0.007(1)$ & $-0.025(1)$ & $0.013(1)$ \\\ns
		\bottomrule
	\end{tabular}
\end{table}

Figure~\figref{fig:reconstructed-LSF-MTF}{b} shows the modulation transfer function of the reconstructed LSF compared to
that of an aberration-free optical system and of the fitted wavefront model.
The MTF of an optical system with a hard aperture has a simple, direct geometrical interpretation, which explains the shape as well as the hard cut-off.
In general, it can be shown that the MTF is directly computed by convolving the pupil function with itself, where  displacements of the electric field distribution in the convolution integral directly translate into spatial frequency units of the MTF~\cite{Goo96}.
Therefore, an optical system with a hard aperture, outside of which the pupil function is strictly zero, results in a cut-off of the MTF at the Abbe frequency.
This cut-off also provides a direct method to extract the actual NA of the optical system without resorting to fitting wavefront distortions.

\begin{figure}[b]
    \centering
    \includegraphics[width=1\columnwidth]{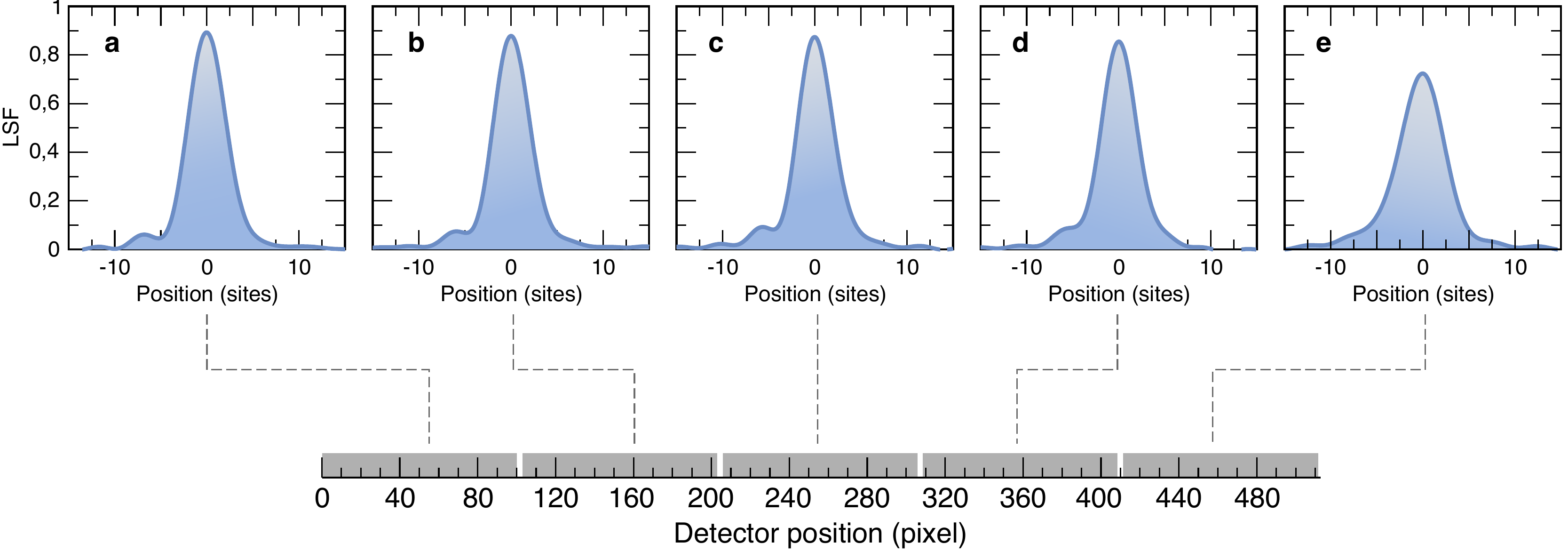}
    \caption{
	Reconstructed LSFs for different patches of the CCD chip. Patches (\textbf{a}-\textbf{c}) show no significant change in the shape of the LSF and can therefore be regarded as an isoplanatic region. The height of the LSF drops for the last patch (\textbf{e}) to below 80\%, whereas the width increases by 16\%. 
  }\label{fig:lsf-position}
\end{figure}

\subsection{Isoplanatic regions}\label{sec:isoplanatic-patch}
\begin{figure}[t]
    \centering
    \includegraphics[width=1\columnwidth]{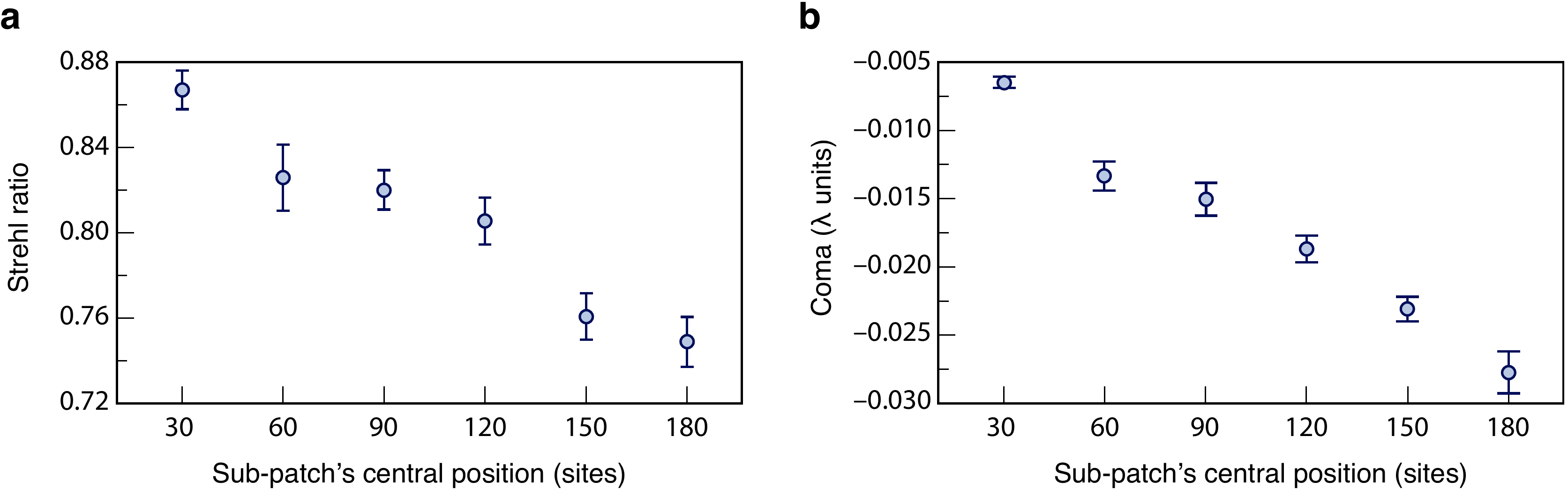}
    \caption{Strehl ratio (\textbf{a}) and coma (\textbf{b}), as an example of spatially varying optical aberration, are shown for six consecutive sub-patches.
  }\label{fig:StrehlRatioPatches}
\end{figure}

The deconvolution problem described in section~\ref{sec:deconvolution-problem} relies on a translationally invariant  response of the optical system.
However, in real systems the LSF varies over the field of view because of optical aberrations primarily due to coma.
Due to spatial variations, the localization precision of the atoms' positions is reduced if a single LSF is employed over the full field of view.
In the literature, regions over which the LSF remains effectively unchanged are known as isoplanatic regions.
To characterize the homogeneity of the LSF of our imaging system, we divide the full CCD region into five patches, each spanning over 100 pixels, where we reconstructed the LSF individually for each patch using the algorithm presented in~\sref{sec:determining-lsf}, see figure~\figref{fig:lsf-position}{a-e}.
The first three patches exhibit minor changes in the optical response, while the rightmost one shows a clearly visible broadening and decreased peak hight.
Fluorescence images of atoms analyzed in this manuscript are from the three leftmost regions only.
To even account for minor spatial deviations, the first three regions are further divided into sub-patches, each of them with an individual local LSF used to reconstruct atoms' position.
The result of the wavefront-deviation analysis of the sub-patches, according to the method presented in the previous section, is illustrated in figure~\ref{fig:StrehlRatioPatches}.

\section{Modeling the noise sources}\label{sec:modelling-noise-sources}
In order to identify the exact lattice-site locations of individual atoms with high reliability, not only the optical response of the imaging system should be precisely known (see previous Sec.~\ref{sec:atomic-sources}) but also an accurate model of all relevant noise sources should be constructed.
{A well-designed imaging apparatus should strive to attain a RMS noise limited by fluorescence photon shot noise, which represents the true fundamental noise contribution.
In general, this can be reached (as is the case of this work) by understanding, and suppressing where required, the technical noise contributions affecting the image formation.
Moreover, one should also be aware of the excess noise adding on top of shot noise, which is caused by the EM register in EMCCD cameras.
This noise contribution effectively decreases the signal-to-noise ratio by a factor $\sqrt{2}$ and cannot be simply eschewed as it is intrinsic to the technology of EMCCD cameras.
Alternative detectors such as CMOS cameras, which also feature small read-out noise, are discussed in Appendix~\ref{sec:detector_comparison}.
In this section, we discuss the relevant noise sources and show that the next noise contribution after fluorescence photon shot noise is photon shot noise caused by background stray light ($\approx \SI{0.5}{\photoelectron/{\sqrtpixel}}$).
}
A summary of individual noise components with their scaling and quantitive estimates is provided in table~\ref{tab:noise-contributions}.

\subsection{Noise sources in the detection process}\label{sec:noise-characterization}

\begin{table}[t]
		\caption{\label{tab:noise-contributions}Noise contributions affecting single-atom imaging, a EMCCD detector cooled to $\SI{-70}{\celsius}$, and $\SI{1}{\second}$ exposure time. The overall noise $\sigma$ is obtained from equation~(\ref{eq:emccd-noise-model}), which takes into account the EM amplification factor $g$ and the excess noise factor ${F}$. RMS noise values extracted from: (a) figure~\ref{fig:signal-noise-log-log}, (b) figure~\ref{fig:clock-induced-charges-analysis}. RMS noise values referring to technical specifications: (c) for inverted mode operation CCD with $T<\SI{-50}{\celsius}$, (d) for EMCCD sensors,  (e) $\SI{10}{\mega\hertz}$ read-out rate.}
	\vspace{3mm}
	\centering 
	\begin{tabular}
		{|m{2.9cm}|m{2.9cm}|m{2.2cm}|l|} \hline Noise type & Physical origin & Scaling & \rms noise \\
		\hline\hline &&&\\[-12pt]
		Fluorescence
		\newline{} photon shot\newline noise & {Uncorrelated scattered
		\newline{} photons} & $\langle S_{\mathrm{fluo}} \rangle^{1/2}$ &$ \sigmaShotnoise^{\;\mathrm{(a)}}\lesssim \SI{4}{\electron}/\!\sqrt{\si{\pixel\second}}$ \\[25pt]\hline &&&\\[-14pt]
		Stray light
		\newline{} shot noise & Spurious\newline{}reflections of\newline{}laser fields
		& $\langle S_{\mathrm{stray}} \rangle^{1/2}$ & $\sigmaStray^{\;\mathrm{(b)}} \sim \SI{0.5}{\electron}/\!\sqrt{\si{\pixel\second}}$  \\[26pt]\hline &&&\\[-14pt]
		Laser intensity\newline{} noise & Technical\newline{}fluctuations & $\langle S_{\mathrm{fluo}}+S_{\mathrm{stray}} \rangle$ & $\sigma_{\mathrm{int}}^{}$  \\[6pt]\hline &&&\\[-14pt]
		Photo response
		\newline non-uniformity & CCD substrate\newline{}inhomogeneity & $\langle S_{\mathrm{fluo}}+S_{\mathrm{stray}}\rangle$ & $\sigma_{\mathrm{PRNU}}^{}$ \\[10pt]\hline &&&\\[-14pt]
		Dark current & Thermal charge\newline{}generation & $\langle S_{\mathrm{therm}} \rangle^{1/2}$ & $\sigma_{\mathrm{therm}}^{\,\mathrm{(c)}}\sim 
		0.01 \textrm{\,--\,} 0.1\,\si{\electron}/\!\sqrt{\si{\pixel\second}}$ \\[6pt]\hline &&&\\[-14pt]
		Clock induced
		\newline{}charges & Charge transfer\newline{} & $\langle S_{\mathrm{CIC}} \rangle^{1/2}$ & $\sigma_{\mathrm{CIC}}^{\,\mathrm{(d)}}\sim 0.05 \textrm{\,--\,} 0.1\,\mathrm{e}^{-}/\!\sqrt{\si{pixel}}$\\[10pt]\hline &&&\\[-14pt]
		Read-out noise & Amplification \&\newline{}digitization\newline{}processes & Read-out rate,
		\newline temperature & $\sigmareadout^{\,\mathrm{(e)}}\sim {50}/M\,\si{\electron}/\!\sqrt{\si{pixel}}$ \\[10pt]\hline &&&\\[-14pt]
		Excess noise & EM amplification & Constant\newline{}factor for \newline{}$g\gg1$ & ${F}=\sqrt{2}$ \\
		\hline 
	\end{tabular}
\end{table}

\begin{description}
\item[Fluorescence photon shot noise] originates from fluctuations in the number $S_{\rm fluo}$ of accumulated photoelectrons.
This noise component originates from three independent stochastic processes: scattering of photons by atoms (Poissonian distributed), photoelectron generation with finite quantum efficiency QE($\lambda_\mathrm{f}$) (binomially distributed), stochastic partitioning by imaging the PSF over several CCD pixels (binomially distributed).
The resulting pixel's fluorescence distribution follows Poissonian statistics characterized by an average fluorescence signal $\langle S_{\mathrm{fluo}}[x_i,y_j]\rangle=I_{\mathrm{fluo}}[x_i,y_j]=\mathrm{QE}(\lambda_\mathrm{f})\,F[x_i,y_j]\,T$, where $F[x_i,y_j]$ is the average photon flux directed onto a given pixel of the CCD sensor, $T$ is the exposure time, and $I_{\mathrm{fluo}}[x_i,y_j]$ the intensity distribution from~equation~(\ref{eq:2d-discrete-imaging-equation}).
The \rms shot noise is ${\sigmaShotnoise[x_i,y_j]={\langle S_{\mathrm{fluo}}}}[x_i,y_j]\rangle^{1/2}$.

\item[Stray light] contributes with Poissonian noise due to fluctuations in the number of photoelectrons $S_{\mathrm{stray}}[x_i,y_j]$.
Stray light can be minimized by shielding the objective and blocking reflections of the molasses laser beams as shown in~\fref{fig:imaging-setup}.
The remaining homogeneous stray-light background yields a \rms noise ${\sigma_{\mathrm{stray}}^{}}=\langle {S}_{\mathrm{stray}}\rangle^{1/2}$.

\item[Illumination intensity noise] is produced by temporal intensity fluctuations of the laser light that illuminates the atoms.
The fluorescence emission rate is directly proportional to the illumination intensity for small saturation parameters (see Sec.~\ref{sec:small-light-signals}).
Hence, fluctuations of the laser intensity result in a \rms noise proportional to the detected signal, $\sigma_{\mathrm{int}}^{}[x_i,y_j]\propto \langle S_{\mathrm{fluo}}[x_i,y_j] \rangle$.

\item[Photo-response non-uniformity (PRNU)] is caused by variations in the pixel geometry and in the substrate material across the CCD chip.
In back-illuminated EMCCD sensors, this also includes the so-called etaloning effect due to interference fringes in the a back-thinned silicon substrate (see App.~\ref{sec:comparing-detectors}).
The \rms noise is proportional to the overall incident photon flux, $\sigma_{\mathrm{PRNU}}[x_i,y_j]\propto \langle S_{\mathrm{fluo}}[x_i,y_j] \rangle + \langle S_{\mathrm{stray}}\rangle $.
Because of its static nature, this noise contribution can be reduced by calibrating the CCD sensor sensitivity with a uniform illumination source in order to remove pixel-to-pixel variations.

\item[Read-out noise] occurs in the charge-to-voltage amplification and analog-to-digital conversion process.
Because this noise component $\sigmareadout$ is not amplified by the EM register, it is effectively suppressed by setting the multiplication gain to large values.

\item[Dark current noise] arises from thermally generated charges.
Buried-channel sensors are affected by two contributions \textemdash bulk and surface dark currents \textemdash  depending whether electron-hole pairs are generated in the depletion region or at the silicon-silicon dioxide interface.
Midgap edge states, either localized in the proximity of bulk impurities or at the front interface, strongly enhance the probability of electrons to thermally hop from the valence to the conduction band through a two-step trap-assisted process~\cite{Balestra:2001}.
A third contribution, in general negligible at low temperatures, comes from diffusion of minority carriers (electrons) from the p-doped silicon substrate into the depletion region.
Because of the large continuum of edge states at the interface with the silicon dioxide layer, the surface contribution dominates by about two orders of magnitude.
Since all dark currents are amplified by the EM register, cooling of the CCD sensor is necessary to detect signals with few photons.
For Peltier-cooled sensors ($T>\SI{-100}{\celsius}$), the contribution by surface dark currents is suppressed in inverted-mode EMCCD chips by applying a large negative bias voltage (multi-pinned phase mode~\cite{Saks:1980}), which creates an inversion layer of holes at the surface preventing electrons from filling the midgap states, and thus suppressing the charge generation process.
Fluctuations of the number of thermally generated electrons in the bulk, $S_\mathrm{therm}$, adds a Poissonian noise component with \rms noise ${\sigma_{\mathrm{therm}}^{}=\langle {S}_{\mathrm{therm}}}\rangle^{1/2}$.

\item[Clock-induced charges] are a spurious electronic signal, $S_\mathrm{CIC}$, generated during the charge transfer process in the CCD sensor when the clock voltage switches the pixel from inversion to non-inversion mode.
The process accelerates holes at the inversion layer back to the heavily doped p-type regions (channel stops), which produce charge carriers through impact ionization.
Despite their ubiquitous presence in every CCD sensor, Clock induced charges (CICs) can only be detected in EMCCD sensors due to the extremely low read-out noise.
CICs are reduced by high parallel transfer rates, small slew rates, and small clock swing~\cite{Janesick:2001}, while they cannot be suppressed by cooling the EMCCD chip (the probability of impact ionization even increases with lower temperatures).
By advanced clock-waveform shaping, modern EMCCD cameras can strongly reduce CICs produced by the vertical transfer process~\cite{Daigle:2010b}.
CICs produced in the serial and multiplication register cannot be simply explained in terms of impact ionization by the accelerated holes.
In modern EMCCD cameras, the generation probability of horizontal CICs per shift step results similar to that of vertical CICs~\cite{Djazovski:2013} in spite of what was originally deemed~\cite{E2V04}.
The stochastic generation of CICs is Poissonian distributed with a \rms noise denoted by $\sigma_\mathrm{CIC}=\sqrt{\langle S_\mathrm{CIC}\rangle }$~\cite{Hirsch:2013}.
\item[Excess noise] arises from the stochastic nature of the gain process in the EM register of EMCCD cameras, which causes an asymmetric broadening of noise distributions.
The resulting noise distribution after amplification by the EM register has an RMS noise increased by the so-called excess noise factor (denoted by ${F}$), which tends to $\sqrt{2}$ for a large number of multiplication stages ($g\gg1$), as can be shown~\cite{Robbins:2003}.

\end{description}

\noindent{}The overall signal measured by the EMCCD camera is the sum of all components,  $S[x_i,y_j]=S_\mathrm{fluo}[x_i,y_j]+S_\mathrm{stray}+S_\mathrm{therm}+S_\mathrm{CIC}$, whose variance $\sigma^2$ is the quadrature sum of all individual noise components,
\begin{eqnarray}
\hspace{-2.5cm}\sigma_{}^{2}[x_i,y_j] &= {F}_{}^{2}\left(\sigmaShotnoise^{2}[x_i,y_j]+\sigma_{\mathrm{stray}}^{2}\hspace{-1pt}+\sigma_{\mathrm{int}}^{2}[x_i,y_j]+\sigma_{\mathrm{PRNU}}^{2}[x_i,y_j]+\sigma_{\mathrm{therm}}^{2}\hspace{-1pt}+\sigma_{\mathrm{CIC}}^{2}\right)+\frac{\sigmareadout^{2}}{g^2} .\label{eq:emccd-noise-model}
\end{eqnarray}
Note that $\sigma^2[x_i,y_j]$ is also the variance of $\epsilon[x_{i}^{}, y_{j}^{}]$, which is defined as $\epsilon[x_{i}^{}, y_{j}^{}]=S[x_i,y_j]-\langle S[x_i,y_j]\rangle$ in equation~(\ref{eq:2d-discrete-imaging-equation}).
Equation~(\ref{eq:emccd-noise-model}) shows that high EM gains $g$ improve the signal-to-noise ratio for signals of few photons by effectively removing the read-out noise component $\sigmareadout$, which in CCD sensors dominates over $\sigmaShotnoise$, instead.
In turn, EMCCD cameras are affected by excess noise through the factor ${F}$, which effectively halves the quantum efficiency.

\subsection{Noise model}\label{sec:noise-model}
The knowledge of each individual noise contribution and their physical scaling with respect to the spatially-dependent fluorescence signal $S_\text{fluo}[x_i,y_j]$ (see Tab.~\ref{tab:noise-contributions}) permits to construct a noise model, which is used in the parametric deconvolution process in section~\ref{sec:deconvolotion-process}.
Based on the functional dependence of $\sigma$ on $S_\text{fluo}$, we rewrite equation~(\ref{eq:emccd-noise-model}) in the form
\begin{equation}\label{eq:noise-model-fit}
    \sigma({\langle S_\text{fluo}^{} \rangle})=\sqrt{\sigma_{\mathrm{b}}^{2} +c_{1}^{2}\,{\langle S_\text{fluo}^{} \rangle}+c_{2}^{2}\,{\langle S_\text{fluo}^{} \rangle}_{}^{2}}\, ,
\end{equation}
where the RMS background noise $\sigma_{\mathrm{b}}$ as well as the coefficients $c_{1}$ and $c_{2}$ are parameters to be experimentally determined.
When the signal $S_\text{fluo}$ is given in photoelectron units, the coefficient $c_{1}$ is directly given by the excess noise factor ${F}=\sqrt{2}$.
The coefficient $c_{2}$ is compatible with zero for our experimental apparatus as shown in the following.

\begin{figure}[t]
    \centering
    \includegraphics[width=0.8\columnwidth]{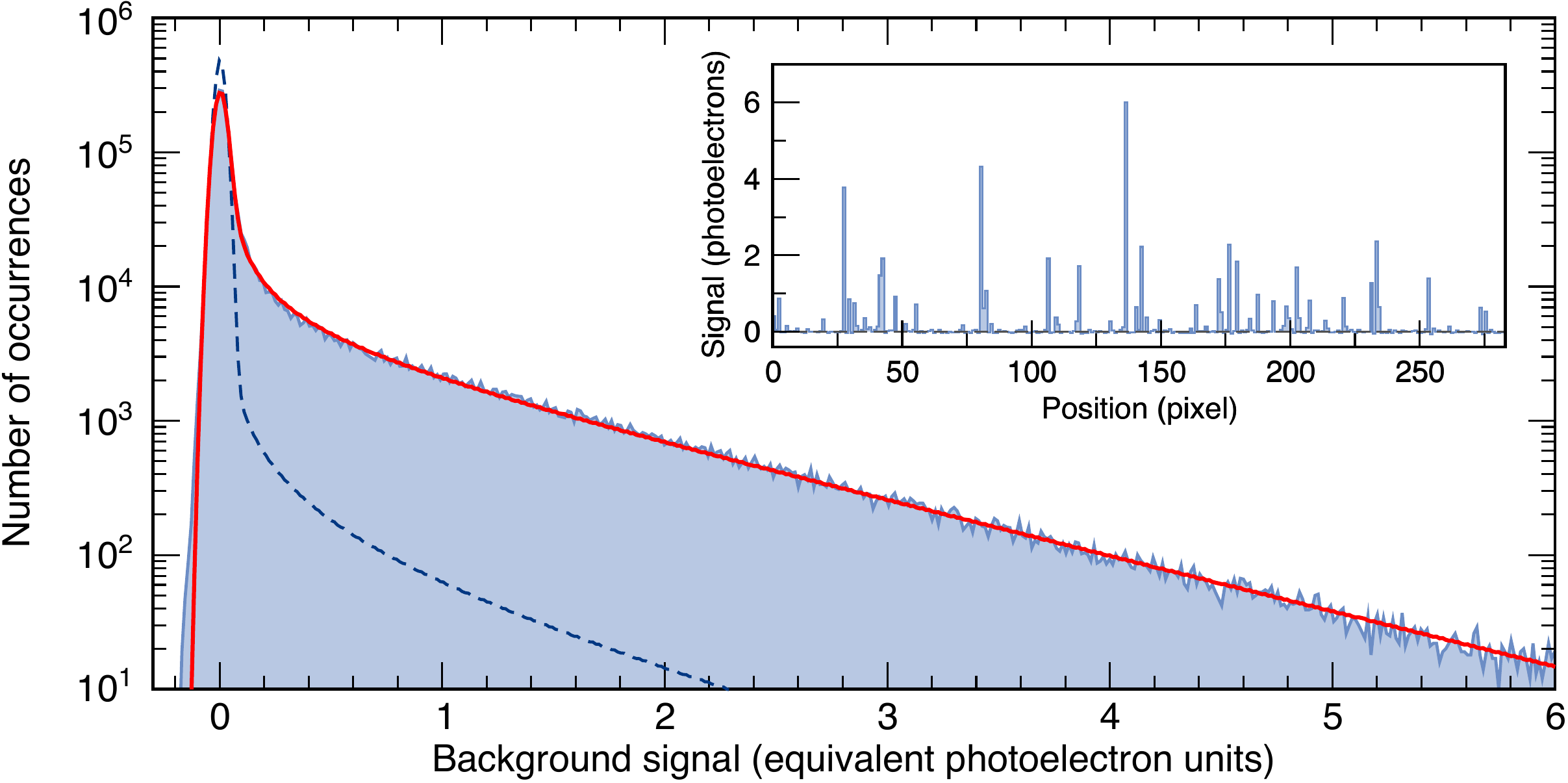}
    \caption{\label{fig:clock-induced-charges-analysis}Histogram of the background noise evaluated on individual CCD pixels. The red solid line depicts a quantitative model fitted to the measured distribution.
The RMS width is dominated by stray light noise ($\SI{0.5}{\photoelectron/{\sqrtpixel}}$), while read-out noise contribute only marginally ($\SI{0.03}{\photoelectron/{\sqrtpixel}}$).
Also shown for comparison is the same model fitted to the background-noise histogram recorded without stray light (blue dashed line).
The inset shows exemplary the signal of a single row of CCD pixels of one of the recorded images, where the spikes originate from CICs.
Note that the histogram contains negative values since the read-out noise is centered at zero.}
\end{figure}

We determine the coefficient $\sigma_{\mathrm{b}}$ by analyzing the background noise of the imaging system from a series of images recorded without atoms in the optical lattice (see inset of figure~\ref{fig:clock-induced-charges-analysis}).
The histogram in figure~\ref{fig:clock-induced-charges-analysis} shows the background signal per pixel binned by signal strength in units of photoelectrons.
The characteristic shape of the histogram mainly arises from read-out noise (Gaussian peak) and stray light (exponential tail).
Calculating the RMS width of the recorded histogram yields the desired coefficient $\sigma_{\mathrm{b}}=\SI{0.6}{\photoelectron/{\sqrtpixel}}$ for $\SI{1}{\second}$ illumination time.
To obtain quantitative insight into the different noise components of the background signal, we model it in terms of Poisson-distributed clock-induced charges, which are stochastically amplified through the EM register, on top of which Gaussian-distributed read-out noise is added~\cite{Lan08}.
The red line in figure~\ref{fig:clock-induced-charges-analysis} shows the fitted model reproducing closely the  recorded background-noise histogram, whereas the dashed blue line shows the same model fitted to a background-noise histogram for images recorded with the camera shutter closed.
Due to the blocked stray light, the exponential tail is reduced by more than one order of magnitude.
However, it does not fully vanish because of dark currents and primarily clock induced charges (see Sec.~\ref{sec:noise-characterization}).
Fitting the background-noise histogram allows not only a more accurate estimate of the RMS background noise ($\sigma_{\mathrm{b}}$), but also a precise calibration of the EM gain $g$, which is a free fitting parameter.
In fact, the parameter $g$ enters the model through the probability to record $y$ electrons after the EM register for $x$ initial electrons \textemdash comprising both spurious electrons and photoelectrons \textemdash which is given by~\cite{Basden:2003}
\begin{equation}\label{eq:prop-EM-amp}
   \mathcal{P}(y|x) = \frac{y^{x-1}\exp(-y/g)}{g^x(x-1)!}\,.
\end{equation}

\begin{figure}
    \centering
    \includegraphics[width=0.65\columnwidth]{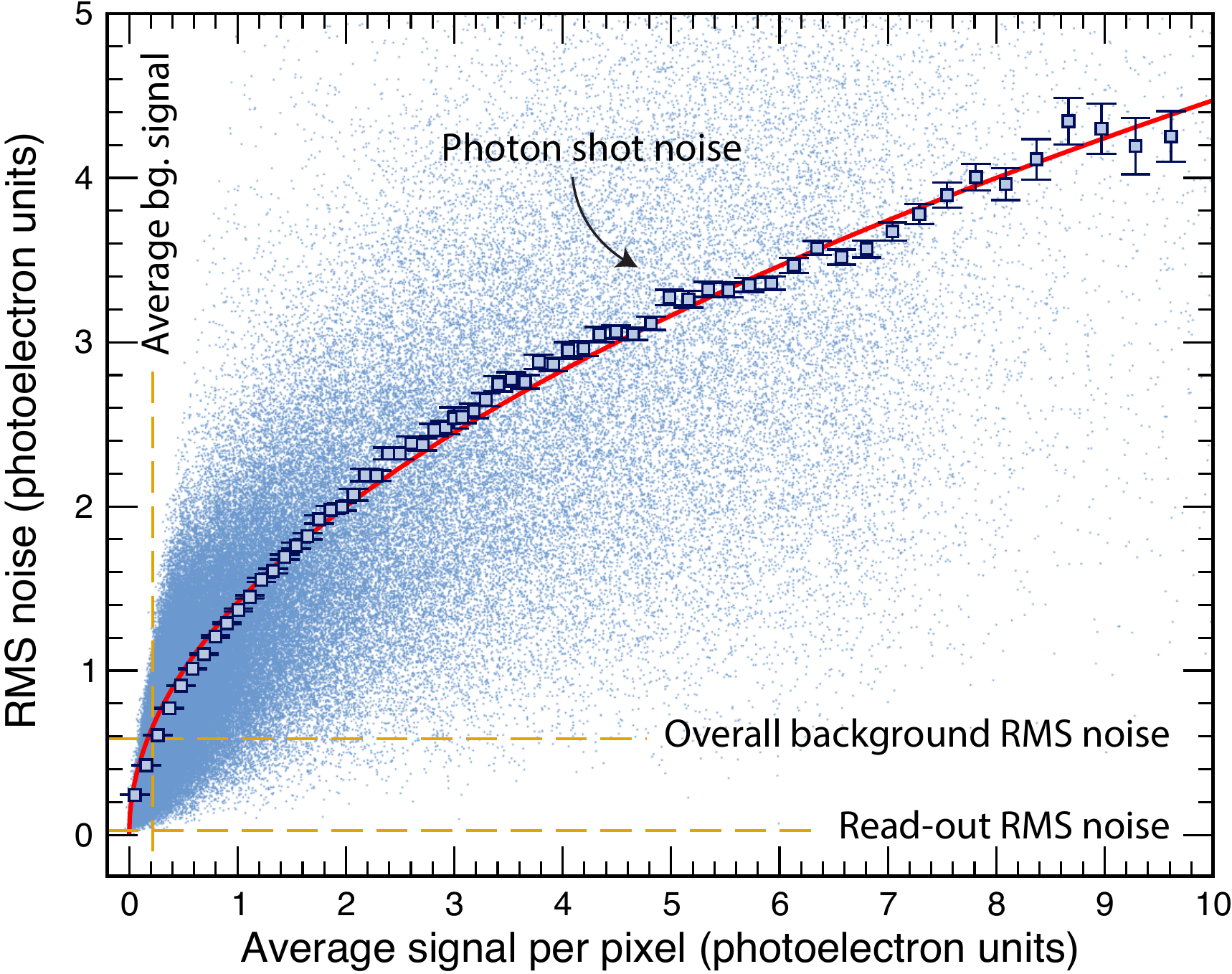}
    \caption{Measurement of the signal-to-noise ratio curve.
	The square points with error bars depict the binned distribution of the estimated signal-to-noise data points (blue dots).
The solid red line is the noise model according to equation~\eqref{eq:noise-model-fit} with no free parameters.
From above, the two horizontal dashed lines indicate the values of $\sigma_\text{b}$ and of the read-out RMS noise, while the vertical dashed line shows the average background signal.}\label{fig:signal-noise-log-log}
\end{figure}

To validate the noise model given in equation~(\ref{eq:noise-model-fit}), we perform a measurement of the signal-to-noise relationship~\cite{Mullikin:1994}.
Based on $>1000$ sets of five consecutive fluorescence images ($\SI{3}{s}$ exposure time each) containing a small ensemble of atoms, we estimate for every CCD pixel the average signal as well as the standard deviation (RMS noise) in each set of images.
Note that we consider only image sets where the distribution of atoms remains constant (neither atom hopping nor losses). 
The cloud of dots in figure~\ref{fig:signal-noise-log-log} shows the correlation between the estimated average signal and the estimated RMS noise, both expressed in photoelectron units.
By binning the signal-to-noise data points by their signal strength, we obtain a precise  reconstruction of the  signal-to-noise relationship, as shown in the figure.
The measurement is in good agreement with the noise model calculated using the coefficients $\sigma_\text{b}$, $c_1$ and $c_2$ given above, therefore confirming the square-root dependence of the RMS noise on the fluorescence signal strength $S_\text{fluo}$.
Furthermore, because no linear dependency is discernible in the recorded signal-to-noise relationship, we set $c_{2}$ to zero.

\section{Localization of atoms by parametric deconvolution}\label{sec:parametric-deconvolution}
We retrieve the position of atoms in the optical lattice using a parametric deconvolution process, which comprises several stages:
(1) The 1D integrated fluorescence images are divided into regions of interest (ROIs), each with a small number of atoms.
(2) The number of atoms is determined for each ROI based on the total number of photoelectrons.
(3) We create a model function of the fluorescence distribution for the given number of atoms and (4) use it to obtain a first estimate of the positions of atoms employing a spectral-density estimation algorithm.
(5) The estimated positions provide the starting values for a non-linear least squares estimate, which yields the location of atoms with improved precision.
(6) We further enhance the localization accuracy by an additional stage that constrains the atoms' positions to the discreteness of the optical lattice and merges all ROIs together.

\subsection{Counting atoms in regions of interest}\label{sec:image-segmentation}

\begin{figure}[b!]
    \centering
    \includegraphics[width=0.7\columnwidth]{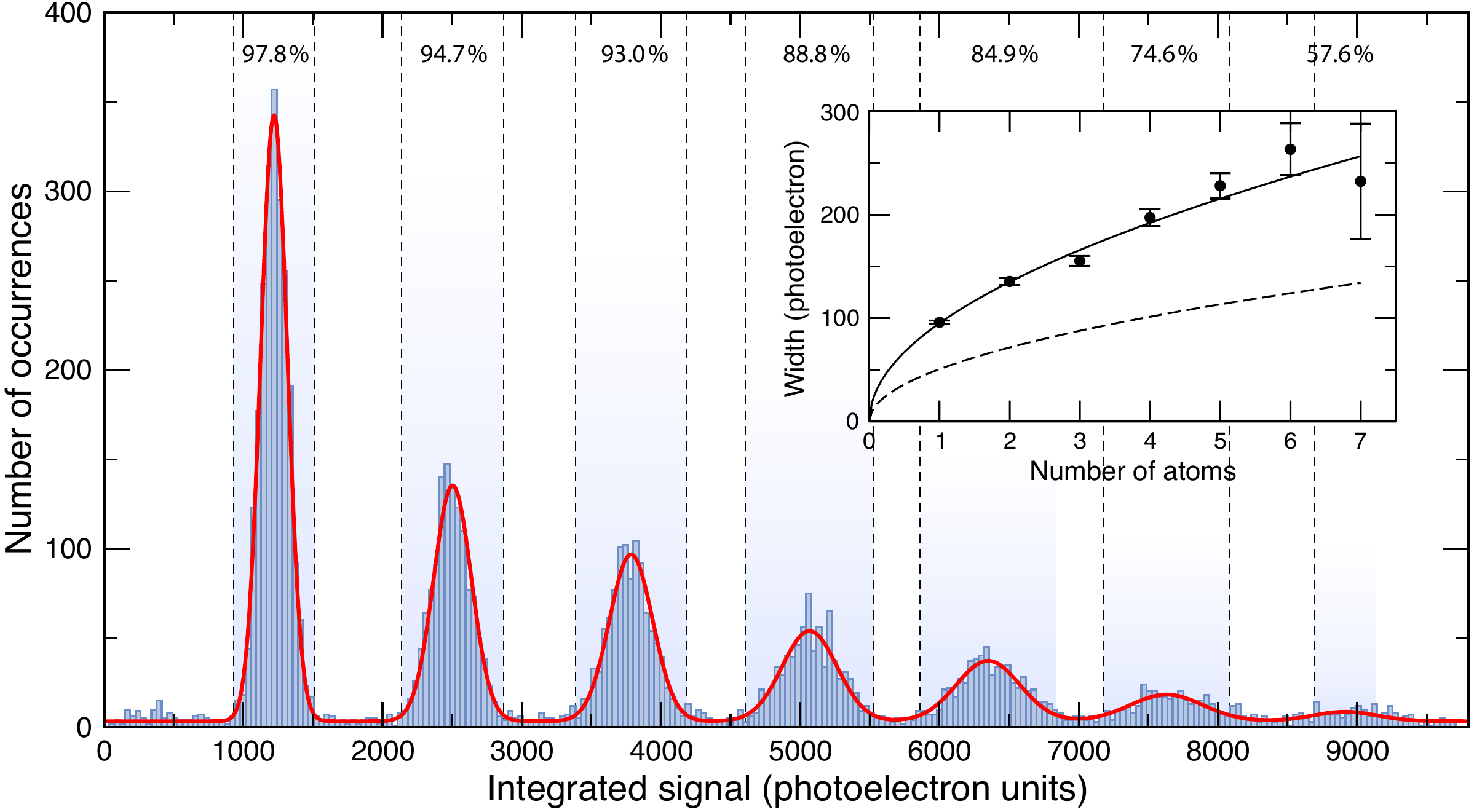}
    \caption{Histogram of integrated photoelectrons from approximately 6000 fluorescence images, each acquired with an exposure time of $1\,\mathrm{s}$.
	The histogram shows equidistant peaks corresponding to different numbers of atoms $m$, whose shape is reproduced by Gaussian functions added to a homogenous background (solid line).
	The height of the peaks is solely determined by the abundance of the corresponding number of atoms in the analyzed dataset.
In the inset: The RMS width of the peaks increases as $\sqrt{m}$ (solid line).
A systematical discrepancy is visible between the measured widths and the curve expected from Poisson statistics of photon counts combined with the EM register's excess noise (dashed line).
We cannot explain this discrepancy in terms of the noise model presented in section~\ref{sec:modelling-noise-sources}.
}\label{fig:total-fluorescence-histogram}
\end{figure}

The knowledge of the exact number of atoms is a necessary prerequisite in order to determine the positions.
While the identification of the number of atoms is relatively straightforward when the atoms are well separated from each other, e.g.\ by counting the number of peaks in the intensity distribution, it is more difficult when the atoms cluster together with separations smaller then the optical resolution.
In such a situation, the peaks of individual atoms are no longer discernible.
Hence, instead of counting peaks, we estimate the number of atoms from the total number of recorded photoelectrons normalized to that of a single atom.
Figure~\ref{fig:total-fluorescence-histogram} shows the histogram of the integrated photoelectrons of a large number of ROIs, exhibiting well-separated equidistant peaks.
Each peak corresponds to a different number of atoms $m$, with the leftmost peak marking the number of photoelectrons per atom (about $\SI{1300}{\photoelectron/\second}$).
The continuous curve in the figure shows the best fit to the recorded histogram based on the sum of seven Gaussian distributions combined with a homogeneous background, which is added to account for atom losses during the exposure time corresponding to the flat background.
The width of the peaks obtained from the fitting model (see inset in figure~\ref{fig:total-fluorescence-histogram}) increases with $\sqrt{m}$.
This broadening implies that adjacent peaks with $m>5$ overlap significantly, thus preventing an unambiguous identification of the exact number of atoms $m$.
This is one reason why we divide every integrated fluorescence distribution into smaller, well-separated ROIs, each containing at least one atom, as shown in figure~\figref{fig:image-segmentation}{b}.
The width of each ROI is determined by thresholding method known as image segmentation algorithm.
Besides, the subdivision in ROIs with a small number of atoms is also beneficial to reduce the computation time of the non-linear last squares estimation of positions, which scales quadratically with the number of atoms in a ROI.

The parametric deconvolution problem requires as a precondition that the number of atoms is correctly determined.
Therefore, it is important to identify acceptance regions $R_i$ of the integrated photoelectron signal where the statistical hypothesis $H_i$ \textemdash the analyzed ROI contains precisely $i$ atoms \textemdash is verified with a probability higher than certain desired confidence levels.
Moreover, we should also take into account the additional statistical hypothesis $H_0$ that one or more atoms are lost during the exposure time.
Referring to the fitting model in figure~\ref{fig:total-fluorescence-histogram}, the \mbox{$i$-th} Gaussian function describes the probability distribution when $H_i$ occurs, while the homogenous background models the probability distribution when $H_0$ occurs.
The acceptance regions $R_i$ (shaded regions in the figure) are obtained by maximizing their width (i.e., the  probability $\mathcal{P}(R_i|H_i)$ defining the power of the statistical test) under the provision that the statistical confidence remains higher than a certain desired confidence level (percentage numbers in the figure).
The confidence level for the region $R_i$ is defined as the probability that the hypothesis $H_i$ is verified when the integrated photoelectron signal is detected in $R_i$, namely $\mathcal{P}(H_i|R_i)$~\cite{StatTest}.
From the analysis of acceptance regions in figure~\ref{fig:total-fluorescence-histogram} we find that, given a certain confidence level, the width of the acceptance region is strongly determined by atom losses (homogeneous background).
This result shows that is beneficial to minimize the atom loss probability during the illumination process ($\sim \SI{1}{\percent}$ in our case).
By post-selection analysis, we only retain ROIs for which the integrated photoelectron signal lies inside one of the acceptance regions. 
Higher confidence levels can be chosen, but this implies narrower regions $R_i$ and, therefore, more ROIs post-selected out.
In order to achieve for the one-atom peak confidence levels $>\SI{99}{\percent}$ with small rejection rates ($<\SI{1}{\percent}$), we segment the fluorescence image in smaller ROIs than those considered in the example displayed in figure~\ref{fig:total-fluorescence-histogram}.

The described method to determine the number of atoms, which relies on the total number of photoelectrons in each ROI, is relatively robust and simple to implement.
However, it does not make optimal use of physical information contained in the image since spatial information is lost after the integrating photoelectrons for each ROI.
It is possible to improve the accuracy by incorporating the spatial information of the recorded fluorescence images through a Bayesian updating algorithm~\cite{Miguel:2016}.

\subsection{The model distribution}\label{sec:modeleing-source-distribution}
When illuminated by nearly resonant light, atoms in a deep optical lattice behave like identical light sources positioned at certain sites of the lattice.
Supposing that the number of atoms  $m$ is exactly known (see Sec.~\ref{sec:image-segmentation}), we model the integrated signal of equation~(\ref{eq:2d-discrete-imaging-equation}) as
\begin{equation}\label{eq:modeled-signal-convolved}
I_\text{M}[x_i]=\sum_{l=1}^{m} A_{l}^{}\hspace{1pt}L_\text{CCD}(x_i^{}-\xi_{l})
\end{equation}
where $\xi_{l}$ are the positions of the atoms, 
and the amplitude factors $A_l$ account for inhomogeneities from the the illumination lasers as well as from atom losses during the exposure time.
$L_\text{CCD}$ is the response function of the imaging system defined in equation~(\ref{eqn:ideal-LSF-reconstruct-convergence}) representing the fluorescence distribution of a single atom with sub-pixel resolution (numerical interpolation between sub-pixels permits its evaluation for any real-valued argument).
Because we perform background subtraction on all integrated intensity distributions, the model in equation~(\ref{eq:modeled-signal-convolved}) does not require an additional constant offset.
In addition, relying on the discreteness of positions in the optical lattice, we can express $\xi_{l}$ as
\begin{equation}
	\label{eq:dfubfivuf}
\xi_{l}=a\hspace{2pt}p_l+\delta_\text{L}\,,
\end{equation}
where $p_l$ are the desired integer positions in lattice-site units, $a$ is the separation between lattice sites in CCD pixel units, and $\delta_\text{L}$ is the position offset of the optical lattice.
For small optical aberrations (see Secs.~\ref{sec:characterizing-the-lsf} and \ref{sec:isoplanatic-patch}), it is sufficient to consider a single value of $a$ ($\num{1.47}$ pixels per lattice site) for the entire field of view.
Moreover, losses by light-induced collisions prevents the detection of two (or more) atoms in the same lattice site in deep optical lattices~\cite{DeP99,Schlosser:2001}.
Hence, $p_l\neq p_{l'}$ for any pair of atoms $l$ and $l'$.

\subsection{The parametric deconvolution process}
\label{sec:deconvolotion-process}
To retrieve the atoms' positions, we employ a non-linear least squares optimization algorithm, which fits the model distribution $I_\text{M}$ in equation~\ref{eq:modeled-signal-convolved} to the recorded fluorescence distributions.
This parametric deconvolution approach allows us to make optimal use of physical information contained in the response function of the imaging system and in the noise model. 
However, non-linear least squares optimizations require well-chosen starting parameters in order to guarantee the convergence to the global optimum.
The parameters of our model are the amplitudes $A_l$ and positions $\xi_{l}$ of atoms.
While an initial estimate of $A_l$ can be directly obtained from the average number of photoelectrons per atom (see Sec.~\ref{sec:image-segmentation}), an estimate of positions $\xi_{l}$ demands a separate procedure.
{Hence, to obtain the first estimate of positions for the non-linear least squares optimization,} we make use of the Wiener deconvolution combined with a spectral density estimation algorithm {(MUSIC algorithm)}.

\paragraph{Wiener deconvolution.}
The main idea underlying our approach to obtain an estimate of $\xi_l$ is understood by considering the Fourier transform of the model distribution in equation~(\ref{eq:modeled-signal-convolved}),
\begin{equation}
	\label{eq:diufifvfvfvkiik}
	\mathcal{F}(I_\text{M})[k]=\text{OTF}_\text{CCD}[k]\sum_{l=1}^{m} A_{l}^{} \hspace{2pt}e^{i\, 2\pi \,k\, \xi_l }= \text{OTF}_\text{CCD}[k]\hspace{2pt}f[k],
\end{equation}
where the convolution theorem enables the optical transfer function $\text{OTF}_\text{CCD}[k]$ (see Sec.~\ref{sec:determining-lsf}) to be factored out of the sum.
The function $f[k]$ is an oscillatory signal containing exactly $m$ Fourier components, whose frequencies are exactly the positions $\xi_l$.
This allows us to recast the problem of estimating the positions $\xi_l$ in that of estimating a discrete number $m$ of frequencies (spectral density estimation).
The presence of noise in the recorded signal $S[x_i]$, however, makes $f[k]$ difficult to be computed from the ratio $\mathcal{F}(S)[k]/\text{OTF}_\text{CCD}[k]$.
In fact, because the noise $\epsilon[x_i]$ has a white spectrum, if we divided the Fourier-transformed recorded signal $\mathcal{F}(S)[k]$ by the reconstructed optical transfer function $\text{OTF}_\text{CCD}[k]$, noise spectral components in the proximity of the Abbe frequency would be strongly amplified owing to the small magnitude of OTF at higher frequencies (see Sec.~\ref{sec:characterizing-the-lsf}).
In order to obtain $f[k]$ avoiding noise amplification, we employ the Wiener deconvolution algorithm \cite{Rajagopalan:2014}, which computes
\begin{equation}
	f[k]=\frac{\mathcal{F}(S)[k]}{\text{OTF}_\text{CCD}[k]}\,\frac{\text{MTF}[k]^2}{\text{MTF}[k]^2+1/\text{SNR}[k]}
\end{equation}
where $\text{SNR}[k]=f[k]^2/\sigma^2$ is the signal-to-noise ratio defined as the ratio between the estimated deconvolved signal $f[k]$, which is obtained by applying the filter iteratively (typically 10 iterations), and the  integrated noise power in the analyzed ROI, which is estimated as $\sigma^2 =n_\perp n_\parallel\, \sigma_b^2+F^2\hspace{1pt}N $ (see also Sec.~\ref{sec:noise-model}). We recall that $n_\perp$ and $n_\parallel$ represent the number of CCD pixels in the ROI in the direction transverse and parallel to the lattice, respectively, and $N$ is the integrated number of photoelectrons.
The Wiener filter factor in equation~(\ref{eq:diufifvfvfvkiik}) is relevant only at higher frequencies, while it is approximately $1$  at lower frequencies because $\text{SNR}[k]$ is very large (typically $>100$) and $\text{MTF}[k]\sim 1$ for $k \, r_\text{A}\ll 1$.

\paragraph{Spectral density estimate (\emph{MUSIC {algorithm}}).}
Several methods exist in the literature to estimate the spectral density from a noisy signal $f[k]$.
The simplest method known as periodogram method employs a discrete Fourier transform of $f[k]$ to determine the $n$ dominant Fourier components, whose frequencies yield an estimate of the positions $\xi_l$.
However, this method suffers from known deficiencies such as being a biased estimator and exhibiting spectral leakage.
{More refined methods known as subspace methods have been developed for the estimation of the spectral components when the signal contains exactly $m$ dominant Fourier components (i.e.\ $m$ atoms) with amplitudes well above the noise background.
Among the subspace methods, the so-called MUSIC algorithm (multiple-signal classification) has been identified as the one exhibiting the highest spectral resolution \cite{Stoica:2005}.
MUSIC yields a pseudospectrum $f[k]$ exhibiting a negligible bias in case of sufficient signal-to-noise ratio \cite{Xu:1992} and not suffering from spectral leakage in contrast to non-parametric methods (e.g., periodogram).
In particular, the strength of MUSIC algorithm for the first estimation of the atoms' position resides in its robustness against noise disturbances and in the fact that no prior knowledge of the parameters (i.e.\ the atoms' positions) is required.
This differs from least-squares minimization procedures, which require good starting parameters to ensure a rapid converge to a global minimum.
While our implementation of MUSIC algorithm requires the prior knowledge of the number of atoms, $m$, in the ROI, extensions of the algorithm exist in the literature that also estimate the number of sources \cite{Wax:1984}, which could be helpful to handle very large ROIs with high filling factors.
}
\begin{figure}
    \centering
    \includegraphics[width=0.9\columnwidth]{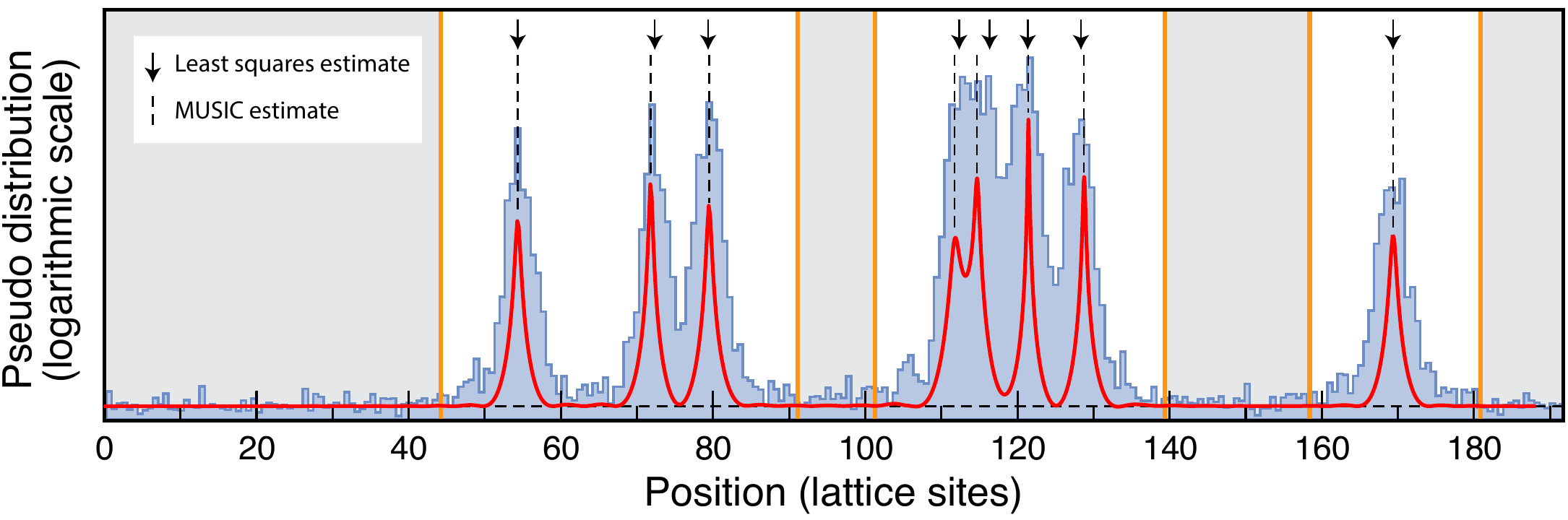}
    \caption{
	Pseudospectrum obtained by the MUSIC spectral density estimate algorithm (in logarithmic scale).
	The narrow peaks provide an estimate of the eight atoms' positions.
	The binned intensity distribution is the same as in figure~\figref{fig:image-segmentation}{b} (in linear scale).
 	The positions estimated by the MUSIC algorithm (vertical dashed lines) are very close within one lattice site to the positions determined by the least squares algorithm (arrows).
	\label{fig:image-segmentation-music}}
\end{figure}
The solid line in figure~\ref{fig:image-segmentation-music} displays the estimated power spectral density obtained from the MUSIC algorithm applied to the fluorescence distribution shown in figure~\ref{fig:image-segmentation}.
The pseudospectrum exhibits eight sharp peaks much narrower than the diffraction-limit width, each approximately centered on the atoms' positions (note the logarithmic scale in the figure). 
The figure shows that the positions estimated by the MUSIC algorithm are very close to those determined by the more accurate non-linear least squares estimator, which takes into account the dependence of noise on the signal. 

\paragraph{Non-linear least squares estimator.}

We use the position of the atoms estimated by the MUSIC algorithm as input parameters of the non-linear least squares minimization 
\begin{equation}\label{eq:least-square-opti}
	\min_{A_{1}\ldots\, A_{n},\,\xi_1\ldots\, \xi_n}\Bigg(\sum_{i=1}^{n_{\hspace{0.3pt}\parallel}}\frac{(S_\text{fluo}[x_i]-I_{\mathrm{M}}[x_i])^2}{\sigma^2(I_{\mathrm{M}}[x_i])}\Bigg)\,
\end{equation}
where $S_\text{fluo}[x_i]$ is the background-subtracted fluorescence distribution, $I_{\mathrm{M}}[x_i]$ the model distribution given in equation~(\ref{eq:modeled-signal-convolved}), $\sigma$ the noise model presented in equation~(\ref{eq:noise-model-fit}), and $n_\parallel$ is the number of pixels in the 1D ROI.
In the minimization problem of equation~(\ref{eq:least-square-opti}), the positions $\xi_l$ are treated as real-valued free parameters (compare Sec.~\ref{sec:enhancing}).
Furthermore, we use the model distribution $I_{\mathrm{M}}$ instead of the measured signal $S$ to estimate the noise variance $\sigma^2$ at the pixel $x_i$ because the model function provides a better estimate of the fluorescence signal after a few iterations of the least squares minimization.
Several algorithms exist to carry out the minimization in equation~\ref{eq:least-square-opti} such as the Levenberg–Marquardt method.
In our case we employ a trust-region algorithm, which allows us to constrain the amplitude parameters $A_l$ to physical boundaries (typically five times the width of the one-atom peak in Fig.~\ref{fig:total-fluorescence-histogram}).
An example of the least squares parametric deconvolution is shown by the solid red line in figure~\figref{fig:image-segmentation}{b}.
{
The accurate model in equation~(\ref{eq:modeled-signal-convolved}) constructed from the measured LSF and the weighting factors in equation~(\ref{eq:least-square-opti}) accounting for the correct variance of noise in each pixel ensure flat residuals with a variance around $1$, as displayed in figure~\figref{fig:image-segmentation}{c}.
For normal distributed residuals, the nonlinear least squares fit is equivalent to a maximum-likelihood estimator of positions \cite{Small:2014}, which defines the gold standard concerning the extraction of physical information from fluorescence images, as argued in section~\ref{sec:intro}.
Because each 1D pixel of the integrated signal carries a large number of fluorescence photoelectrons (see Fig.~\figref{fig:image-segmentation}{b}), the dominating Poisson-distributed shot noise is well approximated by a Gaussian distribution.
However, excess noise in the EMCCD camera causes non-Gaussian deviations, which can be seen, for example, in the logarithmic graph of figure~\ref{fig:clock-induced-charges-analysis}.
Even neglecting this super-Poissonian noise characteristic, previous work using EMCCD cameras reports localization of single emitters with a precision attaining the Cramér-Rao information bound \cite{Smith:2010}.
It is the purpose of future work to refine our estimation of the atoms' positions by maximizing the appropriate likelihood function in order to account for the EM excess noise \cite{Chao:2013} as well as for the Poissonian statistics in the limit of very small signals \cite{Ashida:2016}.
}
{In addition, we find that the} distribution of the sum of squared residuals obtained by analyzing the positions of atoms in $>5000$ ROIs is well described by a $\chi^2$ distribution with $n_\parallel-2m$ degrees of freedom.
{This result suggests that} the minimization procedure of equation~(\ref{eq:least-square-opti}) {approaches the limit of} the maximum-likelihood estimator of the atoms' positions.

\subsection{Enhancing the localization precision at higher filling factors}\label{sec:enhancing}
The parametric deconvolution method outlined in section~\ref{sec:deconvolotion-process} works very reliably in case of ROIs containing only a few atoms separated by several lattice sites.
This is the situation, for example, of single-particle experiments such as quantum walks~\cite{Karski2009a, Genske2013, Robens2015}.
However, the determination of the atoms' positions is less reliable for atoms clustered in small ensembles, where the lattice filling factor approaches unity~\cite{Kar09}.
In experiments investigating strongly interacting particles, it is particularly important to reconstruct the atoms' positions with a high reliability also when the spacing between particles is close to, or is even less than, the optical resolution of the imaging system~\cite{Yamamoto:2015}.
By taking the discreteness of the lattice into account, we demonstrate that the previously presented parametric deconvolution method can be extended to achieve high success rates also for small ensembles of atoms that are closely packed.
As argued in section~\ref{sec:modeleing-source-distribution}, the fact that atoms are trapped in an optical lattice provides us with two pieces of information:
(1) the distance between two atoms can only be a multiple of the intersite separation $a$ (see Eq.~(\ref{eq:dfubfivuf})) and (2) two or more atoms cannot occupy the same lattice site.
To exploit the discreetness of the optical lattice, the lattice constant $a$ (\SI{433}{\nano\meter}) needs to be precisely known in units of CCD pixels.
Its value can directly be computed from the magnification factor (see Sec.~\ref{sec:imaging-setup}) and the pixel size (see Sec.~\ref{the-emccd-detector}) or more accurately measured by analyzing the distribution of distances between two atoms, which are obtained using the deconvolution method presented in the previous section.

\begin{figure}
    \centering
    \includegraphics[width=0.7\columnwidth]{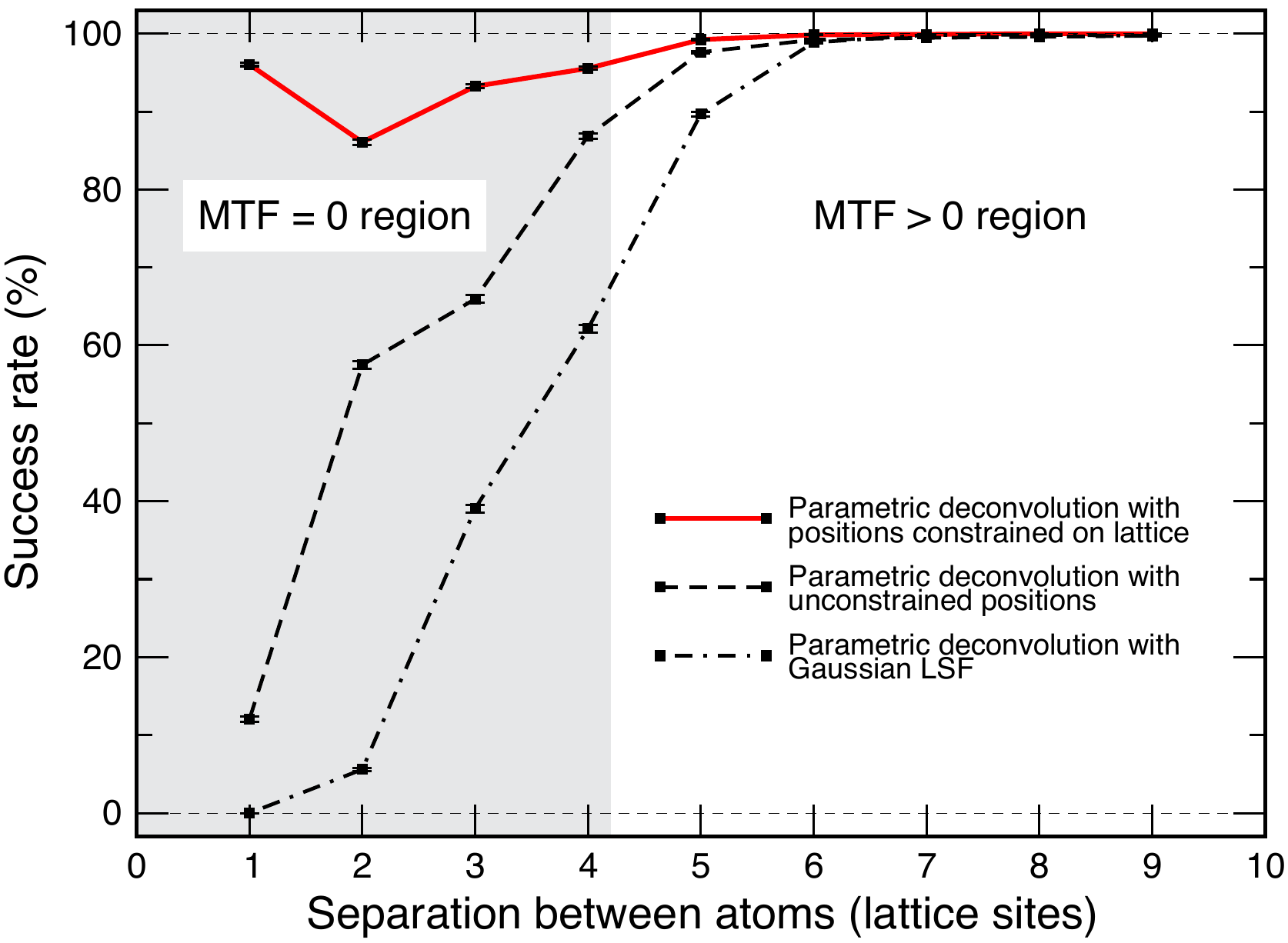}
    \caption{Success rates of finding the correct distances between four atoms {equally spaced} using different parametric deconvolution methods.
	{Note that the success rates refer to the correct identification of all three distances separating the four atoms.}
	For each point, we analyze $\num{10000}$ simulated images of four atoms with equidistant separations ($\SI{1}{\second}$ exposure time).
	The solid line includes the constraints on the atoms' positions by the optical lattice, the dashed line refers to the continuous deconvolution method with unconstrained positions, and the dot-dashed line shows for comparison the  success rate when Gaussian functions are used instead of the reconstructed LSF.
	 All fits produce good success rates for separations larger than the Abbe limit (right-hand side), but only the discrete parametric deconvolution method achieves high success rates for all lattice filling factors. 
	\label{fig:benchmarking}}
\end{figure}

To include the constraints (1) and (2), we adopt the following procedure:
the positions of the atoms are initially determined by the parametric deconvolution process described in the previous section, and rounded to an integer multiple of lattice sites.
We subsequently produce an array of all combinations of distances between atoms, where each distance is let vary by $\pm1$ lattice site with respect to the initial estimate.
Furthermore, we exclude all combinations where two atoms occupy the same lattice site.
For each combination of distances, we perform a non-linear least squares minimization of equation~(\ref{eq:least-square-opti}) with the amplitudes $A_l$ and the lattice position offset $\delta_\text{L}$ as the only free parameters.
We finally choose the combination of distances with the maximum likelihood, which provides us with the best guess of the positions of atoms.
Moreover, the $\chi^2$ distribution of the sum of squared residuals (see Eq.~(\ref{eq:least-square-opti})) allows us to perform a likelihood-ratio test (e.g., Neyman–Pearson lemma) that rejects the reconstructed positions if the statistical confidence lies below a certain specified value.
{A related approach has also been reported in reference~\citenum{She10}, however, without discussing how noise contributions are handled in the deconvolution problem.}

In order to quantitatively benchmark the reliability of the discrete deconvolution method against the continuous one presented in the previous section, we employ Monte Carlo simulations of fluorescence images, which provide large statistics and the exact knowledge of the true positions.
To simulate a pattern of atoms in the lattice, the model distribution in equation~(\ref{eq:modeled-signal-convolved}) is used to construct the fluorescence images, where the noise is randomly drawn to reproduce shot noise fluctuations of the fluorescence signal and the background noise distribution shown in figure~\ref{fig:clock-induced-charges-analysis}.
Our Monte Carlo simulations also incorporate the stochastic fluctuations produced by the EM register.
In particular, we simulated four atoms equally spaced with the spacing varying from one to nine lattice sites.
Figure~\ref{fig:benchmarking} shows the success rate in determining the correct distance between the four atoms.
The analysis of simulated images shows that the success rate of the continuous parametric deconvolution rapidly decreases for separations smaller than the Abbe radius $r_\text{A}$ (diffraction limit).
The drop in the success rate is even more evident when a Gaussian function is used instead of the precisely reconstructed LSF $L_\text{CCD}$, see section~\ref{sec:determining-lsf}.
In contrast, it is remarkable that the success rate of the discrete parametric deconvolution remains above $\SI{90}{\percent}$ for almost all lattice filling factors.
Moreover, the success rate even increases at unity filling since the number of possible configurations of the four atoms is strongly reduced.

{We recently developed a new atom resorting technique that allows us to deterministically position a small ensemble of atoms in any arbitrary pattern on the 1D optical lattice.
The experimental details of the resorting technique are the subject of a future publication~\cite{Robens:2016}.
We employed this technique to reposition, on demand, four atoms along the lattice, thus reproducing the same distributions studied in figure~\ref{fig:benchmarking}, with the atoms separated by an equal number of sites (10, 5, 2 or even 1 lattice sites).
Though based on a significantly smaller statistics, the experimental results are consistent with the theoretical findings obtained with a large number of simulated images, confirming an} enhancement in the reliability of the parametric deconvolution {if the positions are} constrained onto the lattice.

\section{Outlook}
Optical diffraction imposes a stringent limit on the bandwidth of an optical system: physical information contained in the spatial frequency components above the Abbe frequency $1/r_\text{A}$ is not captured by the imaging system.
However, this physical information is not irremediably lost as long as prior knowledge of the structure of the imaged object is available. 
Advances in image processing techniques\textemdash especially driven by the field of super-resolved fluorescence microscopy\textemdash have demonstrated that in this case the higher-frequency components of the imaged object's spectrum can be extrapolated from the imaged distribution.
This principle is what enables information retrieval with spatial resolution beyond the diffraction limit.
The prior knowledge is necessary to solve the deconvolution problem, which ideally reconstructs the original object's distribution (and its entire spectrum) by eliminating diffraction-induced blurring and noise effects.
In this article, we presented state-of-the-art methods that solve the deconvolution problem for fluorescence images of neutral atoms in optical lattices making optimal use of the prior information on the physical system (sub-pixel-reconstructed line spread function, accurate noise model, discreteness of the optical lattice).
The image processing methods we developed are applicable to any experimental apparatus for fluorescence imaging of trapped atoms and ions, and can be directly generalized to two dimensions.

Our methods are particularly beneficial to improve the localization reliability in experiments with constraints on the number of scattered photons, or on the numerical aperture of the objective lens.
For example, experiments imaging light fermions like lithium and potassium atoms suffer from low fluorescence scattering rates, which are generally more than one order of magnitude smaller than those achieved with heavier atoms like Cs and Rb.
Recent experiments with light fermions have shown that even for large numerical apertures $\text{NA}>0.8$ the number of photoelectrons recorded per atom is around $\num{1000}$ for an exposure time of $\SI{1}{\second}$ ~\cite{Haller:2015,Cheuk:2015,Parsons:2015}. 
Similar yields of photoelectrons are obtained in our apparatus where we employ an objective lens with much smaller numerical aperture ($\text{NA}=0.23$).
By taking advantage of our deconvolution methods, these experiments can minimize the exposure time while ensuring a high reliability to localize each individual atom in the lattice.
Short exposure times reduce the probability of atoms to hop to neighboring sites, thus avoiding localization errors as well as losses of atoms colliding inelastically with a second neighboring atom.

In our laboratory, the construction of a {new} experimental apparatus for imaging single atoms {with much higher numerical aperture ($\text{NA}>0.9$)} is underway, which ensures a twenty times higher collection efficiency and a four times narrower point spread function.
{The analysis methods demonstrated in this article, when applied to the new imaging apparatus,} should enable single-site resolution with {unprecedentedly short exposure times} ($<\SI{10}{\milli\second}$) \cite{NewObjective}, {allowing us} to directly discriminate between the two internal hyperfine states of Cs atoms (qubit states)~\cite{Gibbons:2011,Fuhrmanek:2011,Miguel:2016}.
Moreover, it is the goal of future work to investigate whether compressed sensing techniques, which rely on a completely different principle than our parametric deconvolution method, provide advantages for information retrieval beyond the diffraction limit~\cite{Szameit:2012,Tang:2013}.

\ack
The authors gratefully acknowledge financial support by the DFG (FOR635) and by the European Community (projects DQSIM and SIQS).
M.~Karski, R.~Reimann, and C.~Robens acknowledge partial support from the Studienstiftung des deutschen Volkes, R.~Reimann, C.~Robens, S.~Brakhane from the BCGS Graduate School, and A.~Alberti from the Alexander von Humboldt Foundation.
The authors also acknowledge L.~F{\"o}rster and J.-M.
Choi for technical assistance, and A.~Steffen and J.~Zopes for valuable discussions on numerical methods, and G.~L\"uchters for insightful discussions on statistics{, as well as Y.~Ashida and C.~Monroe for stimulating and fruitful discussions on the limits of super-resolution imaging.}

\section*{References}
\bibliographystyle{njpbibstyle}
\bibliography{imaging-paper}

\begin{thebibliography}{10}
\bibitem{StatTest}
{The conditional probability can be computed as \protect{$\mathcal{P}(H_i|R_i)
  = \mathcal{P}(H_i)\hspace{1pt}\mathcal{P}(R_i|H_i)/\hspace{-1pt}\sum_j
  \mathcal{P}(H_j)\hspace{1pt} \mathcal{P}(R_i|H_j)$}.}
\bibitem{NewObjective}
{Because the localization precision in
  Eq.~\ref{eqn:localozation-precision} scales as $\Delta x\propto
  \lambda_f/(\text{NA}\sqrt{N})$ (assuming only shot noise), the twenty times
  larger collection solid angle and the four times narrower point spread
  function should enable single-site resolution with exposure times around
  \protect{$\SI{1000}{\milli\second}/(16\cdot 20) \approx
  \SI{3}{\milli\second}$}.}
\end{thebibliography}

\begin{appendices}
\section{Single-photon cameras}\label{sec:detector_comparison}
With current technology, the sensitivity limit of conventional CCD sensors is determined by read-out noise, which is produced during the processes of charge-to-voltage and analog-to-digital conversion~\cite{Seitz:2011}.
Low-noise CCD cameras cooled to low temperatures ($<\SI{-50}{\celsius}$) and operating at high read-out rates (\mbox{$>\SI[per-mode=symbol]{10}{\mega\pixel\per\second}$}) have \rms values of read-out noise equivalent to $\SI{6}{\photoelectron}$.
When operating at low read-out rates ($<\SI{20}{\kilo\hertz}$), commercial state-of-the-art CCD sensors can attain lower read-out noise around $\SI{2}{\electron}$, however, still above the value of one electron per pixel \textemdash the ultimate limit for single photon imaging.
Recently, research prototypes showed that sub-electron read-out noise could be realized in the near future~\cite{Fernandez2012}.

For weak radiation sources (e.g.\ fluorescing single atoms) the amount of shot noise can amount to very few electrons.
One would ideally need CCD sensors with sub-electron sensitivity to avoid that the read-out noise dominates over shot noise.
To overcome the technical limit imposed by read-out noise, three major technologies have been developed over the years and found widespread application: electron multiplying CCD cameras and intensified CCD cameras, and more recently CMOS sensors.
All these technologies rely on the preamplification of the physical signal \textemdash the number of photoelectrons \textemdash prior to the read-out amplification stage.
These three technologies are shortly reviewed in the following.

\subsection{Intensified CCD cameras.}
The basic idea underlying intensified CCD (ICCD) cameras dates back to the first half of the 20th century: It consists in employing a photocathode to convert the incoming photons into photoelectrons, whose number can then be multiplied by avalanche amplification.
A gating electric field is used to precisely control the access of photoelectrons into a microchannel plate, where the avalanche multiplication process takes place.
The electrons exiting the microchannel plate are accelerated towards a phosphor screen, upon which they recreate the same distribution of photons impinging at the photocathode.
Secondary photons emitted by the phosphor screen are eventually imaged onto a low-noise CCD detector.
Electron multiplication in the microchannel plate can readily reach amplification factors up to $\num{e4}$, which allow read-out noise to be effectively suppressed down to values as low as $\SI[per-mode=symbol]{e-3}{\electron\per\pixel}$.
For a more detailed account of ICCD cameras, please refer to Ref.~\citenum{Nuetzel:2011} and references therein.

\subsection{Electron multiplying CCD cameras.}
The first practical demonstration of electron multiplying CCD (EMCCD) cameras has been provided in 2001~\cite{Jerram:2001,Hynecek:2001,Mackay:2001}.
In essence, EMCCD sensors are CCD sensors equipped with a low-noise electron multiplying (EM) register in addition to the conventional register used to transport electron charges.
In comparison to conventional CCD registers, the EM register uses a higher clocking voltage to provide the electrons with sufficient kinetic energy to cause impact ionization.
Through avalanche multiplication, the EM register is able to stochastically multiply the number of electrons by factors in the range of a few thousands.
The read-out noise is thereby effectively reduced on the order of $\SI[per-mode=symbol]{e-2}{\electron\per\pixel}$, which is smaller than noise produced by dark counts and clock induced charges (CIC)~\cite{Ives:2009}.
For a more detailed account, please refer to Ref.~\cite{Robbins:2011} and references therein.
In addition, both CIC and read-out noise can be further suppressed by hardware binning the CCD pixels along the vertical direction~\cite{Westra:2009}.
It is understood that spatial resolution along the binned direction is reduced depending on how many pixels are vertically binned together.
Note that hardware binning is not exploited in this work.

\subsection{Comparison between ICCD and EMCCD cameras}
In contrast to EMCCD sensors, ICCD sensors are virtually insensitive to spurious CIC and thermal dark electrons since these processes occur after the amplification process.
They are, however, sensitive to dark electrons generated in the photocathode (so-called equivalent background illumination), whose rate is generally small $\ll\SI[per-mode=symbol]{1}{\electron\per\pixel\per\second}$ already at room temperature.
Therefore, cooling of the sensor to low temperatures is not needed for short exposure times lasting only a few seconds.
On the other hand, ICCD cameras have a lower quantum efficiency than EMCCD cameras (see also discussion in App.~\ref{sec:comparing-detectors}) because of the lower sensitivity of photocathode materials, especially at longer wavelengths.
For instance, at our fluorescence wavelength $\lambda_\text{f}=\SI{852}{\nano\meter}$, $\mathrm{QE}(\lambda_{\mathrm{f}})$ does not exceed $\SI{20}{\percent}$ at the present time.
ICCDs allow much shorter gate times than EMCCDs, though this feature is not particularly relevant for imaging single atoms trapped in optical potentials.
In addition, the finite radius of the microchannel plate in ICCD sensors limits the resolution to about 50 line pairs/mm; for the Abbe radius $r_\mathrm{A}$ of our microscope objective, for instance, a magnification factor of the order of 50 is required to avoid affecting the overall optical resolution.
A detailed comparison of noise properties has been published elsewhere~\cite{2003Denvir,2003coates,Dus04}.

\subsection{CMOS sensors}
Due to significant advances over the past two decades in manufacturing microscale, ultralow-noise MOSFET devices, CMOS image sensors represent today an appealing alternative to conventional CCD detectors in low light imaging applications~\cite{2008Theuwissen,Zimmermann2009}.
The basic element consists here of an active pixel sensor, which provides the charge-to-voltage conversion electronics and the transistors needed for voltage buffering and pixel addressing~\cite{Magnan:2003}.
The absence of the CCD shift register enables faster parallel read-out rates and excludes the noise contribution caused by CIC.
Read-out noise in CCD sensors is dominated by Johnson noise at the charge amplifier, whose white spectrum is inevitably fed into the large video bandwidth.
In contrast, parallel amplification in CMOS sensors makes it possible to directly amplify the signal at the active pixel location, where the signal is formed.
In such a way, the bandwidth of the source-follower and column amplifier used to amplify the charge signal into a voltage signal can be limited through a low-pass filter.
This prevents high-frequency noise components to be amplified and fed through the high-bandwidth analog-to-digital conversion circuitry~\cite{Kozlowski1999,Janesick:2002b}.
Commercially available CMOS cameras exhibit read-out noise as low as $\SI[per-mode=symbol]{1}{\electron\per\pixel}$, which is nearly independent of the video frame rate.
The presence of transistors in the pixel area significantly screen the silicon photosensitive area from the impinging photons, thus reducing the pixel's fill factor.
To circumvent this problem, an array of microlenses is usually employed in scientific-grade sensors to efficiently collect photons in front of each pixel.
Alternatively, back-illuminated CMOS sensors (see e.g.\ OmniVision Technologies, Inc.) can also be employed.
A detailed comparison of CMOS cameras with CCD, ICCD, and EMCCD cameras has been carried out elsewhere~\cite{2007hain,Krishnaswami2014}.

\section{Front vs.\ back illuminated CCD detectors}\label{sec:comparing-detectors}

In silicon-based detectors, the quantum efficiency (QE) of CCD detectors is determined by the probability that an incoming photon is converted into an electron-hole pair inside the photosensitive region \textemdash a region that is completely depleted of mobile charge carriers \textemdash where electrons are efficiently collected into the pixel by means of a built-in electric field.
For CCD detectors that are frontally illuminated, photons must first transverse the polycrystalline silicon structure of electrodes and a silicon-oxide insulating layer before they can reach the photosensitive region.
Reflections and absorptions by the electrodes cause a reduction of QE, which can be avoided if the back side, i.e. the one opposite to the CCD electrodes, is turned towards the radiation source.
This can be achieved by etching the chip to a thin layer around $10-\SI{20}{\micron}$ thick.
Back-illuminated back-thinned CCD detectors have thereby doubled the QE with peak values above 
$\SI{90}{\percent}$.

\begin{figure}[t]
    \centering
    \includegraphics[width=0.5\columnwidth]{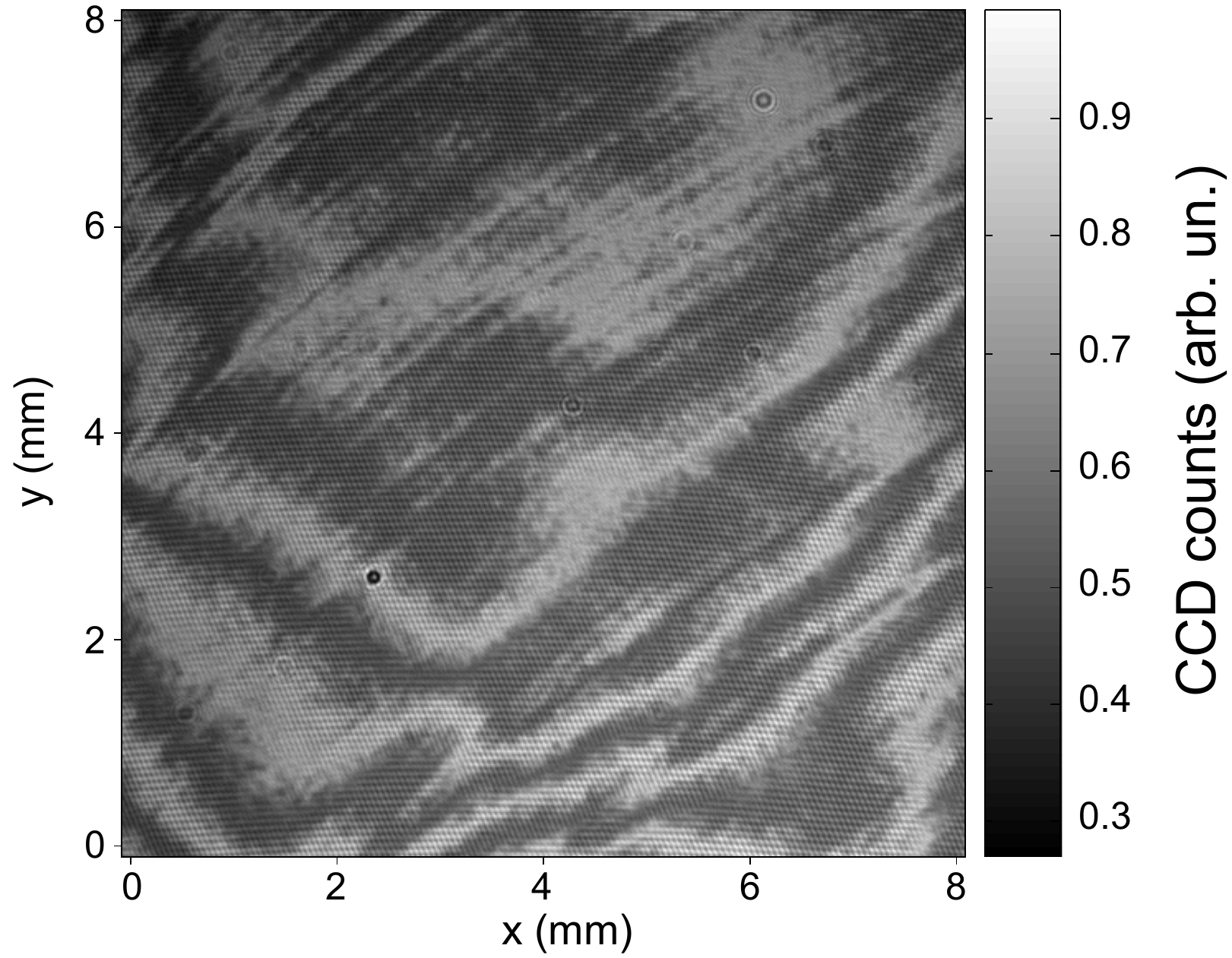}
    \caption{Interference fringes in a back illuminated chip under coherent illumination at $\lambda_\mathrm{f}$. The average peak-to-valley variation amounts to about 40\% of the signal. A peak-to-valley fringe corresponds to a thickness change of the back-thinned silicon layer by less than 100 nm. The axes shows the spatial scale corresponding to the CCD chip.}\label{fig:histogram+fringes}
\end{figure}

We have tested a commercial, state-of-the-art back-illuminated EMCCD sensor with read-out noise specified at ${<\SI[per-mode=symbol]{0.05}{\electron\per\pixel}}$ and dark current noise at ${<\SI[per-mode=symbol]{5e-3}{\electron\per\pixel\per\second}}$ for a temperature below $\SI{-65}{\celsius}$.
This sensor provides a QE of $\SI{60}{\percent}$ at the fluorescence wavelengths of $\SI{852}{\nano\meter}$, which is about the double of that achieved by our front-illuminated camera.
As a downside, the sensor displays interference fringes caused by multiple reflections of monochromatic photons at the interface between the substrate and the silicon-oxide layer and at the interface between the substrate and the air.
In fact, the silicon substrate behave like an etalon plate at longer wavelengths where the absorption depth of silicon increases.
The etaloning effect is evidenced in figure~\ref{fig:histogram+fringes}, where variations of the substrate thickness results in visible fringes with a contrast $\sim \SI{40}{\percent}$.
We have verified that the interference pattern is stable and has only a weak dependence on temperature, as the interference pattern shifted by about half a fringe for a temperature change of $\SI{20}{\celsius}$.
Despite the large intensity variations across the sensor, the stability of interference pattern allows us to filter it out by dividing the recorded signal by a calibrated spatial mask.

Larger reverse-bias voltages ($\sim\SI{100}{\volt}$) together with thicker substrates ($\sim \SI{100}{\micron}$) of high-resistivity silicon results in much wider photosensitive regions (deep depletion)~\cite{2006holland}.
Thicker depletion regions permit to enhance the QE especially in the near-infrared, where the absorption depth of silicon is ${>}\SI{10}{\micron}$.
Because of the higher thickness, photons are converted into electron-hole pairs before reaching the silicon-oxide layer, with the result that the etaloning effect is strongly suppressed.
However, the suppressed etaloning effect comes along with about a one hundred-fold increase of dark counts, which can be suppressed by cooling to low temperatures $<\SI{-50}{\celsius}$.
Up to the present, there exists neither CMOS nor EMCCD cameras that are based on a deep-depletion back-thinned sensor, which would be ideal for detecting small signals in the near-infrared.

\section{Asymptotic limit of the iterative point-spread-function reconstruction}\label{sec:math-convergence-lsf}
We study the LSF produced by the iterative reconstruction algorithm in the limit of infinitely many single-atom images that are overlapped with sub-pixel resolution according to the algorithm presented in section~\ref{sec:determining-lsf}. 
Although the following discussion specifically focuses on the reconstruction of LSF in one dimension, analogous results can be demonstrated for the PSF in two dimensions.

The algorithm first constructs from the image of an atom positioned in $x_0$ a new real-valued distribution with sub-pixel resolution,
\begin{equation}
	\label{eq:fdibvifbvrfro}
I_{x_0}(x)  = \!\!\sum_{n=-\infty}^\infty ( L_\text{CCD}(n\,\Delta_\text{s}-x_0) + \epsilon[n\,\Delta_\text{s}]) \:\,\mathcal{R}_\text{sp}(x-n\,\Delta_\text{s})
\end{equation}
where $L_\text{CCD}$ is the response of the imaging system defined in equation~(\ref{eqn:ideal-LSF-reconstruct-convergence}), $\epsilon$ is the additive noise, $\Delta_\text{s}$ is the spacing between CCD pixels, and $\mathcal{R}_\text{sp}(x)$ is the sub-pixel function (one for $-\Delta_\text{sp}/2<x<\Delta_\text{sp}/2$ and zero elsewhere, where $\Delta_\text{sp}=\Delta_\text{s}/s$ for some integer $s$).
In essence, equation~(\ref{eq:fdibvifbvrfro}) maps the signal $L_\text{CCD}(n\,\Delta_\text{s}-x_0)$ recorded at the $n$-th CCD pixel to a $s$-fold narrower signal $\mathcal{R}_\text{sp}(x-n\,\Delta_\text{s})$, yielding a continuous distribution in the real-valued variable $x$.
The function $L^{(k-1)}_\text{guess}$, which is produced by the algorithm at the iteration $k-1$, is fitted to the recorded fluorescence distribution to  provide a maximum-likelihood estimator $\tilde{x}_0$ of the atom position $x_0$.
This estimator is a stochastic variable due to the noise term $\epsilon$ in equation~(\ref{eq:fdibvifbvrfro}).
Because of symmetry reasons, its probability distribution is symmetrically centered on $x_0$ (unbiased estimator).
Subsequently, the reconstruction algorithm translates with sub-pixel resolution the distribution in equation~(\ref{eq:fdibvifbvrfro}) by $-\tilde{x}_0$, producing a new distribution $I_{x_0}[x+\tilde{x}_0]$.
The algorithm adds all repositioned single-atom intensity distributions to yield a new estimate of the LSF
\begin{equation}
			\label{eq:fhvifbviff}
	L^{(k)}_\text{guess}=\int_{-\infty}^\infty\!\!\mathrm{d}x_0\;\mathcal{P}(x_0)\!\int_{-\infty}^\infty\!\!\mathrm{d}\tilde{x}_0\;\mathcal{P}(\tilde{x}_0|x_0)\,I_{x_0}(x+\tilde{x}_0) 
\end{equation}
where $\mathcal{P}(x_0)$ is the probability of the single atom to be in $x_0$, and $\mathcal{P}(\tilde{x}_0| x_0)$ is the conditional probability expressing the uncertainty distribution of the maximum-likelihood estimator $\tilde{x}_0$.
In the asymptotic limit of infinitely many images, the noise contribution $\epsilon$ averages out so that the expression of equation~(\ref{eq:fhvifbviff}) takes the form
\begin{equation}
	\label{eq:fuivbifbfvfofo}
		\hspace{-15mm}L^{(k)}_\text{guess}=\int_{-\infty}^\infty\!\!\mathrm{d}x_0\;\mathcal{P}(x_0)\!\int_{-\infty}^\infty\!\!\mathrm{d}\tilde{x}_0\hspace{-2pt}\sum_{n=-\infty}^\infty \mathcal{P}(\tilde{x}_0|x_0)\, L_\text{CCD}(n\,\Delta_\text{s}-x_0)\,\mathcal{R}_\text{sp}(x+\tilde{x}_0-n\,\Delta_\text{s})\,.
\end{equation}
This expression  can be further simplified by assuming an isoplanatic response function of the imaging system such that  $\mathcal{P}(\tilde{x}_0| x_0) = \mathcal{R}_x(x_0-\tilde{x}_0)$, where $\mathcal{R}_x$ is the uncertainty distribution of the maximum-likelihood estimator of the atom's position.
In addition, we make the physical assumption that $\mathcal{P}(x_0)$ is uniformly distributed, which is ensured by 
the incommensurability of the optical lattice with respect to the CCD array and by small drifts in the time of the optical lattice.
The latter condition is particularly important to guarantee that all sub-pixels of the reconstructed LSF are equally sampled.
%
After some algebra, equation~(\ref{eq:fuivbifbfvfofo}) can be recast in the form
\begin{equation}
	L^{(k)}_\text{guess}(x)=(\mathcal{R}_x\ast\mathcal{R}_\text{sp}\ast L_\text{CCD})(x)\,,
\end{equation}
which proves the expression used in equation~(\ref{eqn:LSF-reconstruct-convergence}).

\end{appendices}

\end{document}